\documentclass[12pt]{iopart}

\usepackage{graphicx}

\begin{document}

\title[PIC simulation of ion beam driven instabilities]{Particle simulation 
study of electron heating by counterstreaming ion beams ahead of supernova
remnant shocks}

\author{M E Dieckmann$^{1}$, A Bret$^{2,3}$, G Sarri$^{1}$, E Perez Alvaro$^{3}$,
I Kourakis$^{1}$ and M Borghesi$^{1}$}
\address{1. Queen's University Belfast, Ctr Plasma Phys, Belfast BT7 1NN, UK}
\address{2. Harvard-Smithsonian Center for Astrophysics, Cambridge, MA 02138, USA}
\address{3. ETSI Industriales, Universidad de Castilla-La Mancha, 13071 Ciudad Real, 
Spain and Instituto de Investigaciones Energeticas y Aplicaciones Industriales, 
Campus Universitario de Ciudad Real, 13071 Ciudad Real, Spain}

\begin{abstract}
The growth and saturation of Buneman-type instabilities is examined with a particle-in-cell 
(PIC) simulation for parameters that are representative for the foreshock region of fast 
supernova remnant (SNR) shocks. A dense ion beam and the electrons correspond to the upstream 
plasma and a fast ion beam to the shock-reflected ions. The purpose of the 2D simulation is 
to identify the nonlinear saturation mechanisms, the electron heating and potential secondary 
instabilities that arise from anisotropic electron heating and result in the growth of magnetic 
fields. We confirm that the instabilities between both ion beams and the electrons saturate by 
the formation of phase space holes by the beam-aligned modes. The slower oblique modes accelerate 
some electrons, but they can not heat up the electrons significantly before they are trapped by 
the faster beam-aligned modes. Two circular electron velocity distributions develop, which are 
centred around the velocity of each ion beam. They develop due to the scattering of the 
electrons by the electrostatic wave potentials. The growth of magnetic fields is 
observed, but their amplitude remains low.
\end{abstract}

\maketitle

\section{Introduction}

The thermalisation of shock-reflected ion beams plays an important role in solar system 
and astrophysical collision-less plasma. Energetic beams, which consist of the ions that 
were reflected by a plasma shock, outrun the shock and interact with the upstream plasma. 
This upstream plasma is the interstellar medium (ISM) for supernova remnant (SNR) shocks 
and the solar wind for most solar system shocks. The ion beam is a source of free energy, 
which is released through wave instabilities. Such instabilities have been observed ahead 
of solar system shocks \cite{Foreshock}. They are also important mechanisms for particle 
acceleration at supernova remnant (SNR) shocks [2-9].

The Buneman instability \cite{Buneman1,Buneman2} develops between one ion beam and one 
electron beam of equal density, which are both spatially uniform, unmagnetized and drift 
relative to each other. Such a plasma is not current neutral. A plasma with no net charge 
and no net current can be composed of two counter-streaming ion beams and one electron 
species, if the total charge density of the ions is that of the electrons and if the 
partial currents of the three beams cancel out each other. This can be an appropriate 
description of the plasma upstream of a shock. One ion beam is composed of the shock-reflected 
ions and the second ion beam and the electrons are provided by the upstream plasma, into 
which the shock expands. Such distributions are observed ahead of the Earth's bow shock 
\cite{Foreshock}. We refer with Buneman-type instability (BTI) to an electrostatic instability, 
which involves an ion beam with a density below that of the electrons.

A BTI develops, if the drift speed between the ion beam and the electrons exceeds the electron 
thermal speed. Thermal damping effects due to the ions can be neglected, as long as they are 
reasonably cold. If the shock-reflected ion beam is much faster than the electron thermal speed 
and sufficiently dense, then its current has to be cancelled out by a high drift speed between 
the background electrons and ions. This drift speed may exceed the electron thermal speed and 
result in a second BTI. The speed of SNR shocks is a few per cent of the light speed c, which 
exceeds by far the electron thermal speed in the ISM. We may expect that the shock-reflected 
ion beam and the counterstreaming ion beam are both sufficiently fast in the electron rest 
frame to render the plasma unstable. Such plasmas have been examined widely in the past with 
particle-in-cell (PIC) simulations \cite{Dawson}. Previous studies have addressed unmagnetized 
\cite{MyUnM,ShimUnM,Pavan} and magnetized plasmas \cite{MyMag, ShimMag} with one-dimensional
PIC simulations, which can not capture the multi-dimensional nature of the wave fields. More 
recently, the BTI has been examined with two-dimensional PIC simulations in unmagnetized and 
magnetized plasma. The electrostatic simulation in Ref. \cite{Amano} has modeled the interaction 
of one fast ion beam with cool electrons. References \cite{Ohira,DieckNJP} investigated ion 
beams that stream with nonrelativistic and mildly relativistic speeds across an orthogonal 
magnetic field. 

We examine here with an electromagnetic 2D PIC simulation how a plasma thermalises, which 
consists of two counterstreaming ion beams with unequal densities that move through an  
unmagnetized electron plasma. The drift speeds between the electrons and each of the ion beams 
exceed the initial electron thermal speed by factors of 3.4 and 17 and two BTI's develop. The 
faster beam has a speed that is comparable to the ones modeled with PIC simulations in Ref. 
\cite{Amano}. The waves driven by both BTI's interact nonlinearly with the electrons, once their 
amplitude is adequate. Trapping by an electrostatic wave accelerates electrons along the wave 
vector and a thermal anisotropy in the electron's velocity distribution develops as we show here. 
It is the purpose of our study to assess, if a thermal anisotropy-driven Weibel instability (TAWI) 
[21-29] is triggered by anisotropic electron heating. Such a secondary instability can not be 
resolved by the electrostatic PIC code used in Ref. \cite{Amano}.

The initial conditions, which we consider here, are representative for a volume element in the 
foreshock, that is so small that the flow speed of the upstream plasma and of the shock-reflected 
ion beam are constant inside. However, ion beam-driven instabilities can also be triggered by the 
return current \cite{Pohl} of the cosmic rays. Similar instabilities may thus occur in a much 
larger volume ahead of the shock. The TAWI could provide \cite{Schlickeiser} the seed magnetic 
field, which speeds up the cosmic ray-driven instabilities that magnetize SNR shocks \cite{Volk,Cosmo}. 
The inital magnetic amplitude is set to zero in our simulation, which simplifies the detection and 
interpretation of magnetic fields that are driven by the BTI through the subsequent TAWI. This
choice can be justified by the weak magnetic fields ahead of SNR shocks, which yield an electron
cyclotron frequency well below the BTI's frequency in the electron rest frame. At the same time,
the magnetic field close to SNR shocks is thought to be strong enough to reflect enough ions to
form a dense beam, provided that the shock is perpendicular.

Our results are as follows. The interaction of the background electrons with the slow and dense 
ion beam from the background plasma yields the faster-growing instability, as we expect from the 
solution of the linear dispersion relation. The electrostatic waves grow and saturate through the 
formation of electron phase space tubes [34-39].
The spectrum of the unstable waves is not unidirectional \cite{Amano} and the phase space tubes are 
spatially bounded. Their extent orthogonal to their wave vector is a few
times their wave length. Trapping accelerates the electrons into the direction of the propagating 
waves. The electron thermal energy along the beam direction reaches twice the value of the electron 
thermal energy orthogonal to the beam direction and this anisotropy results in the growth of magnetic 
fields. The magnetic fields do, however, only reach an energy density that is 2-3 orders of magnitude
below that of the electric fields and their strength does not exceed that of the interstellar magnetic
fields.

The electrons are initially accelerated by trapping to a speed, which is about twice as high as the 
background ion speed in the electron rest frame. After this initial acceleration, the electrons are 
scattered by the waves. The reflections are elastic, because the electrostatic wave potential is 
almost time-stationary in the reference frame of the bulk ions, and they change the momentum 
direction of the electrons. Repeated collisions between the electrons and the phase space tubes 
redistribute the electrons into a circular interval in velocity space, which is centered at the 
phase speed of the waves. This mechanism differs subtly from that proposed in Ref. \cite{Amano}, 
which has attributed the crescent formation to the low phase speed of the oblique modes. 

The plasma is no longer free of current after the saturation of the first BTI. The electrons 
have been decelerated in the rest frame of the background ions by their interaction with the waves
while the ions with their high inertia have only been weakly affected. 
A macroscopic beam-aligned electrostatic field grows in response to this current on scales much 
larger than the wave length of the BTI \cite{Experiment}. This field accelerates the electrons 
nonresonantly such that the mean electron speed remains close to the one prior to the wave 
saturation, thereby reducing the net current. 

Eventually the electrons start to interact with the faster ion beam. Electrons are accelerated 
primarily by trapping, but the Landau damping of the oblique modes results in a separate population 
of hot electrons. A crescent distribution develops, but it is less pronounced than the one found in 
Ref. \cite{Amano}. The electron heating by the first BTI and the electron acceleration by the 
macroscopic electric field have changed the electron distribution function to a non-Maxwellian 
one and they have decreased the ratio between beam speed and electron thermal speed to a value 
well below that used in Ref. \cite{Amano}, which may explain the different plasma evolution. The 
hotter ion beam we consider here will also modify the wave spectrum of the BTI and the nonlinear 
response of the beam ions to the electric fields, which may also contribute to the different
interaction of the electrons with the waves driven by the second BTI. An energetic circular 
distribution in velocity space gradually develops after the second BTI has saturated by electron 
trapping. The cause is again the electron scattering by the electrostatic structures that move 
now with the mean velocity of the fast ion beam. No TAWI develops in this case during the 
simulation time, even though the electrons continue to show a thermal anisotropy. The absence of 
any significant magnetic field growth during the growth and saturation of the two BTIs suggests 
that the thermal anisotropy created by the nonlinear interaction between electrostatic waves and 
electrons is not capable of magnetising the foreshock of supernova remnant shocks, at least not 
for our initial conditions. 

The structure of the paper is as follows. Section 2 discusses the initial conditions and the PIC 
simulation code. Section 3 presents the simulation results. Section 4 is the discussion. 

\section{The initial conditions and the particle-in-cell simulation code}

\subsection{Initial conditions}

\begin{figure*}[t]
\includegraphics[width=0.41\columnwidth]{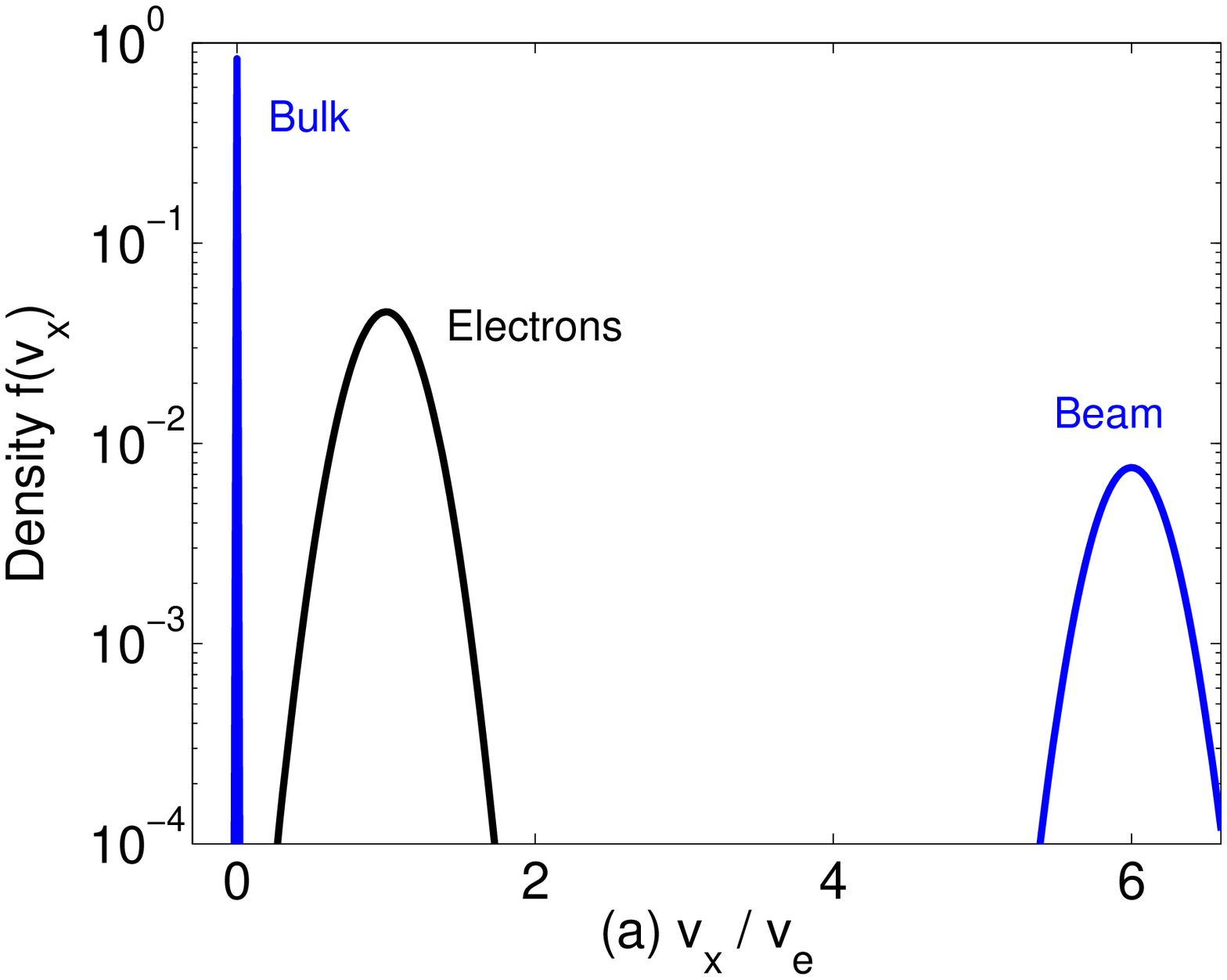}
\includegraphics[width=0.58\columnwidth]{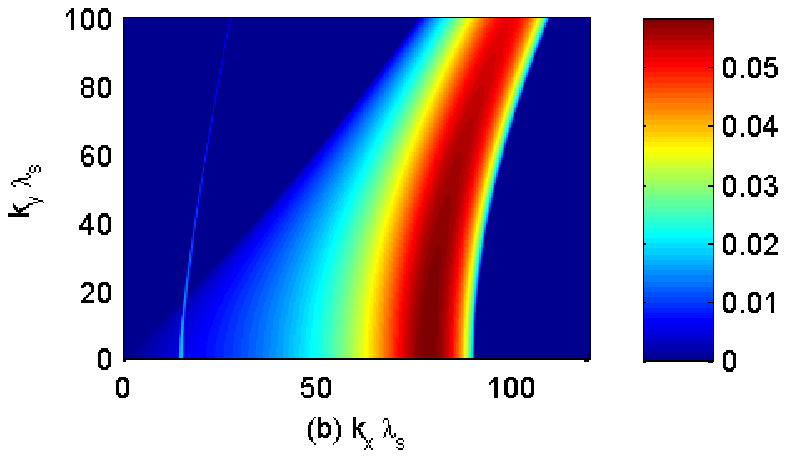}
\caption{(Color online) The beam distribution is displayed in panel (a). Panel (b) shows the 
numerical solution of the linear dispersion relation as a function of the beam aligned wave 
number $k_x c/\omega_p$ and the perpendicular wave number $k_y c/\omega_p$. The color corresponds
to the exponential growth rate of the wave, expressed in units of the electron plasma frequency
$\omega_p$. The unstable wave branch at large $k_x$ arises from the interaction between background 
ions and electrons and that at low $k_x$ from the interaction between electrons and beam ions.}
\label{Fig1}
\end{figure*}

We consider here three spatially uniform plasma species with non-relativistic Maxwellian velocity 
distributions. The electrons have the density $n_0$, the plasma frequency $\omega_p ={(n_0 e^2 / m_e 
\epsilon_0)}^{1/2}$ and the temperature 10 eV or the thermal speed $v_{te} = 1.325 \times 10^6$ m/s. 
They move at the speed $v_e = 4.5 \times 10^6$ m/s along increasing values of $x$, which gives 
$v_e = 3.4 v_{te}$. We consider for computational reasons a reduced mass ratio between the ions 
and electrons of $m_i / m_e = 10^3$. The bulk ions are at rest in the simulation box. Their 
density and temperature are $n_i=5n_0/6$ and 5.5 eV. The bulk ions would correspond to the 
relatively cool upstream plasma ahead of a SNR shock. The ion beam has the density $n_b = n_0 / 6$ 
and temperature 10 keV. Its mean speed is $v_b = 6 v_e$. The ion beam is much hotter than the 
bulk ions. Plasma shocks are neither perfectly planar nor elastic reflectors, so the momentum 
change by the reflection differs for each ion \cite{Matsu}, which causes a rise of the temperature. 
The velocity distribution of the three species is displayed in Fig. \ref{Fig1}(a). The plasma is 
charge neutral by $n_b + n_i = n_0$ and it is initially free of current by $n_0 v_e = n_b v_b$. 

Both BTI's result in electrostatic waves that are practically stationary in the rest frame of the 
respective ion beam. The real parts of the wave frequencies $\omega_{u1}, \omega_{u2}$ and the 
corresponding wave numbers $k_{u1},k_{u2}$ along $x$ can be estimated in the cold plasma 
approximation with the dispersion relation $1-\omega_{b}^2/\omega^2 - \omega_p^2 /{(\omega -k_x 
v_0)}^2=0$ valid for one ion beam with plasma frequency $\omega_b$ and the electrons with the drift 
speed $v_0$. The bulk ions and the electrons drive waves with the frequency $\omega_{u1} \ll 1$ and 
the wave number $k_{u1}c/\omega_p \approx c/v_e$ or $k_{u1} c / \omega_p \approx 67$. Since the 
reference frame of the bulk ions is that of the simulation box, this wave will be practically 
stationary in the latter. 

The instability between the fast ion beam and the electron beam drives 
waves with $k_{u2}c/\omega_p \approx c /(v_b-v_e)$ or $k_{u2}c/\omega_p \approx 13$. Their Doppler 
shifted frequency is $\omega_{u2} \approx k_{u2}v_b$ or $\omega_{u2} \approx 1.2$ in the box frame 
of reference. The growth rate of a BTI is $\omega_i \approx {({3\sqrt{3} \omega_b^2 \omega_p}/
16)}^{1/3}$ \cite{MyUnM} and the growth rate ratio between both BTI's is thus $\omega_{i1} /\omega_{i2} 
= {(n_i / n_b)}^{1/3}$. The exponential growth rate of the BTI driven by the bulk ions is 1.8 times 
that of the BTI driven by the fast ion beam. 

In the cold plasma limit, this growth rate is independent of the wavenumber $k_y$ orthogonal 
to the beam velocity vector. A more accurate growth rate map is obtained from the solution of 
the linear dispersion relation that takes into account thermal effects and is valid for waves 
in the simulation plane with $k^2 = k_x^2+k_y^2$,
\begin{equation}
\left ( \omega^2 \epsilon_{xx} - k_y^2 c^2 \right )\left ( \omega^2 \epsilon_{yy} - k_x^2 c^2 \right )
- {\left ( \omega^2 \epsilon_{yx} + k_x k_y c^2 \right )}^2 = 0,
\end{equation}
and with the elements of the dielectric tensor
\begin{equation}
\epsilon_{\alpha \beta} (\mathbf{k},\omega) = \delta_{\alpha \beta} + \sum_j \frac{\omega_{pj}^2}
{\omega^2}\int v_\alpha \frac{\partial f_j^0}{\partial v_\beta}  d^3v +
\sum_j \frac{\omega_{pj}^2}{\omega^2}\int v_\alpha v_\beta \frac{\mathbf{k}\cdot \left 
(\frac{\partial f_j^0}{\partial \mathbf{v}} \right ) }{\omega -\mathbf{k} \cdot \mathbf{v}} d^3 v,
\end{equation}
where $\omega_{pj}$ and $f_j^0$ are the plasma frequency and equilibrium distribution function
of the $j^{th}$ species, respectively (See section 2 in \cite{Review}). We approximate $f_j^0$ by 
waterbag distributions with the same densities and drift velocities as the plasma species we 
introduce in our simulation. The thermal width of each waterbag distribution equals the thermal 
speed of the corresponding Maxwellian distribution. The waterbag distribution is equivalent to a 
warm fluid model, and a good approximation to a Maxwellian, if the beam speed is large compared
to the thermal spreads \cite{FluidWater}. The solution of the linear dispersion relation is shown 
in Fig. \ref{Fig1}(b), which displays the imaginary part of the wave frequency (growth rate) as a function of the 
beam-aligned wave number $k_x$ and of the perpendicular wave number $k_y$. The colour corresponds to 
the growth rate, expressed in units of the electron plasma frequency $\omega_p$. The 
most unstable wave numbers in this growth rate map match the ones from the cold plasma approximation 
for $k_y \approx 0$. Larger values of $k_y$ result in increasing values of $k_x$ and in lower growth 
rates.

\subsection{The simulation code and the diagnostics}

Our 2D3V electromagnetic and relativistic PIC simulation code \cite{Eastwood} resolves the $x,y$ 
plane and the particle positions in this plane. It updates all three components of the magnetic 
field $\mathbf{B}$, of the electric field $\mathbf{E}$, of the current $\mathbf{J}$ and of the 
relativistic particle momentum $\mathbf{p}$. The code preserves $\nabla \cdot \mathbf{E} = \rho 
/ \epsilon_0$ and $\nabla \cdot \mathbf{B} = 0$ to round-off precision and evolves the 
electromagnetic fields in time through
\begin{equation}
\nabla \times \mathbf{E} = -\frac{\partial \mathbf{B}}{\partial t}\, , \, \,
\nabla \times \mathbf{B} = \mu_0 \epsilon_0 \frac{\partial \mathbf{E}}
{\partial t} + \mu_0 \mathbf{J}.
\end{equation}
 
An ensemble of computational particles (CPs) is followed in time. Each CP with the index $i$ of the 
species $j$ in the simulation has the mass and charge $m_j$ and $q_j$, the momentum $\mathbf{p}_i = 
m_j \Gamma_i \mathbf{v}_i$ and the position $\mathbf{x}_i = (x_i,y_i)$. Each CP corresponds to a 
volume element of the phase space distribution of the particle species it represents and the ratio 
$q_j / m_j$ must be equal to that of the corresponding physical particle. The position and momentum 
of each CP are updated according to
\begin{equation}
\frac{d \mathbf{x}_i}{dt} = \mathbf{v}_i \, , \,
\frac{d \mathbf{p}_i}{dt} = q_j \left ( \mathbf{E}[\mathbf{x}_i] + \mathbf{v}_i \times 
\mathbf{B}[\mathbf{x}_i] \right ) \label{eq1}.
\end{equation}
The simulation box has the size $L_x \times L_y = 3c/\omega_p \times 4c/\omega_p$, which is resolved by 
a grid with $N_x \times N_y = 580 \times 760$ quadratic cells with the sidelength $\Delta_x \approx 1.2 
\lambda_D$, where $\lambda_D$ is the electron Debye length. The boundary conditions are periodic, which 
limits the wavenumber spectrum to $k_x c/\omega_p = 2\pi a / 3$ and $k_yc/\omega_p = 2\pi b / 4$ with the 
integers $1 \le a \le N_x / 2$ and $1 \le b \le N_y / 2$. The short unstable wave has $k_{u1}L_x \approx 
200$ and about 32 wave lengths are resolved. The long unstable wave has $k_{u2} L_x \approx 39$ and about 
6 wave periods are resolved. Each of the plasma species is represented by 320 CPs per cell. The simulation 
time $T_M \omega_p = 1200$ is resolved by the time steps $\Delta_t \omega_p = 2.5 \times 10^{-3}$.

We compute the following quantities to monitor the plasma evolution. The electric field energies $D_{ex}(t) 
= {(2\epsilon_0)}^{-1} \int E_x^2 (x,y,t) dx \, dy$ and $D_{ey}(t) = {(2\epsilon_0)}^{-1} \int E_y^2 (x,y,t) 
dx \, dy$ show the time-evolution of the BTI's. The energy 
$D_{bz}(t) = {(2\mu_0)}^{-1} \int B_z^2 (x,y,t) dx \, dy$ reveals if an electron thermal anisotropy 
$A \neq 0$ with $A=\langle v_x^2 \rangle / \langle v_y^2 \rangle -1$, $\langle v_x^2 \rangle = \sum_s 
{(v_{x,s}-v_e)}^2$ and $\langle v_y^2 \rangle = \sum_s v_{y,s}^2$ results in a magnetic field growth. The 
summation is here over all computational electrons. The electron mean speed along x and y varies in the 
simulation only by about $10^{-2}v_{te}$ and can thus be taken to be constant. The kinetic energies of 
the electrons and of the bulk ions are computed in our relativistic code as $K_e (t) = m_E c^2 \sum_s 
(\Gamma_s - 1)$ and $K_i (t) = m_I c^2 \sum_s (\Gamma_s - 1)$, where $m_E$ and $m_I$ are the non-relativistic 
masses of the computational electrons and ions. The sum is over all CPs of the respective species and 
$\Gamma_s$ the Lorentz factor of the $s^{th}$ CP. All energies are computed in the rest frame of the bulk 
ions and they are normalized by $K_e (0) = E_{K0}$. The mean electric field in the simulation box is 
$\langle E_x \rangle = {(N_x N_y)}^{-1} \sum_{i=1}^{Ny} \sum_{j=1}^{Nx} E_x (i,j,t)$, where $E_x (i,j,t)$ 
is the electric field at the grid cell with the indices $i$ and $j$ at the time $t$. A low pass filter 
removes the oscillations with a frequency exceeding $\omega_f =0.8 \omega_p$, which reveals the trend of 
the curve. 

In what follows, the velocities $\mathbf{v}$ and positions $\mathbf{x}$ are normalized to $v_e$ and to 
the electron skin depth $\lambda_s = c / \omega_p$. The time is multiplied with $\omega_p$. Frequencies 
are given in units of $\omega_p$ and the wave numbers are given as $k c / \omega_p$. The electric and 
magnetic fields are expressed as $e \mathbf{E}/ \omega_p c m_e$ and $e\mathbf{B} / \omega_p m_e$.

\section{Numerical simulation results}

Electrostatic waves are polarized parallel to their wave vector $\mathbf{k}$. Their spectrum 
contains oblique modes and $E_x$ and $E_y$ will both grow. 
\begin{figure*}[t]
\includegraphics[width=0.49\columnwidth]{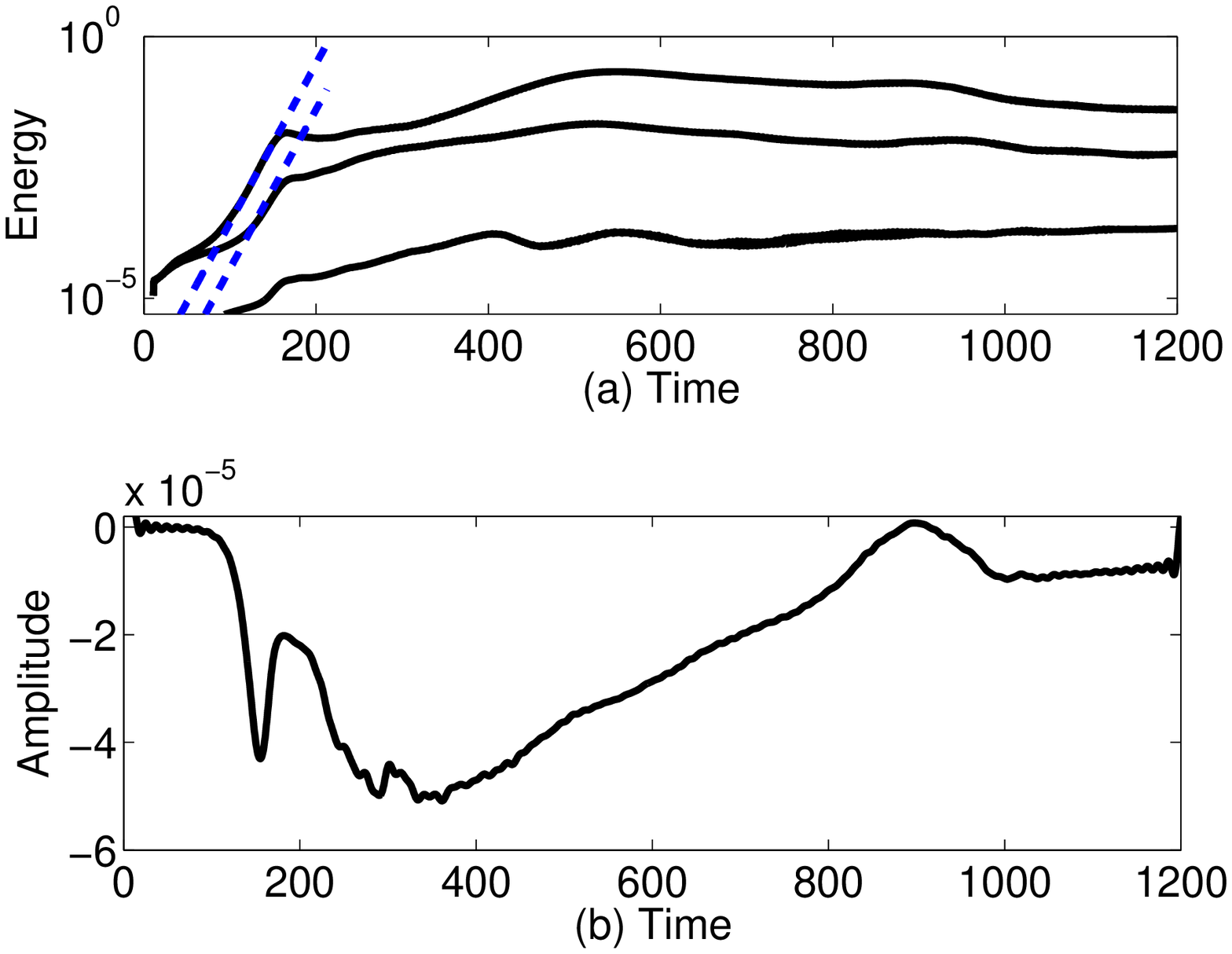}
\includegraphics[width=0.49\columnwidth]{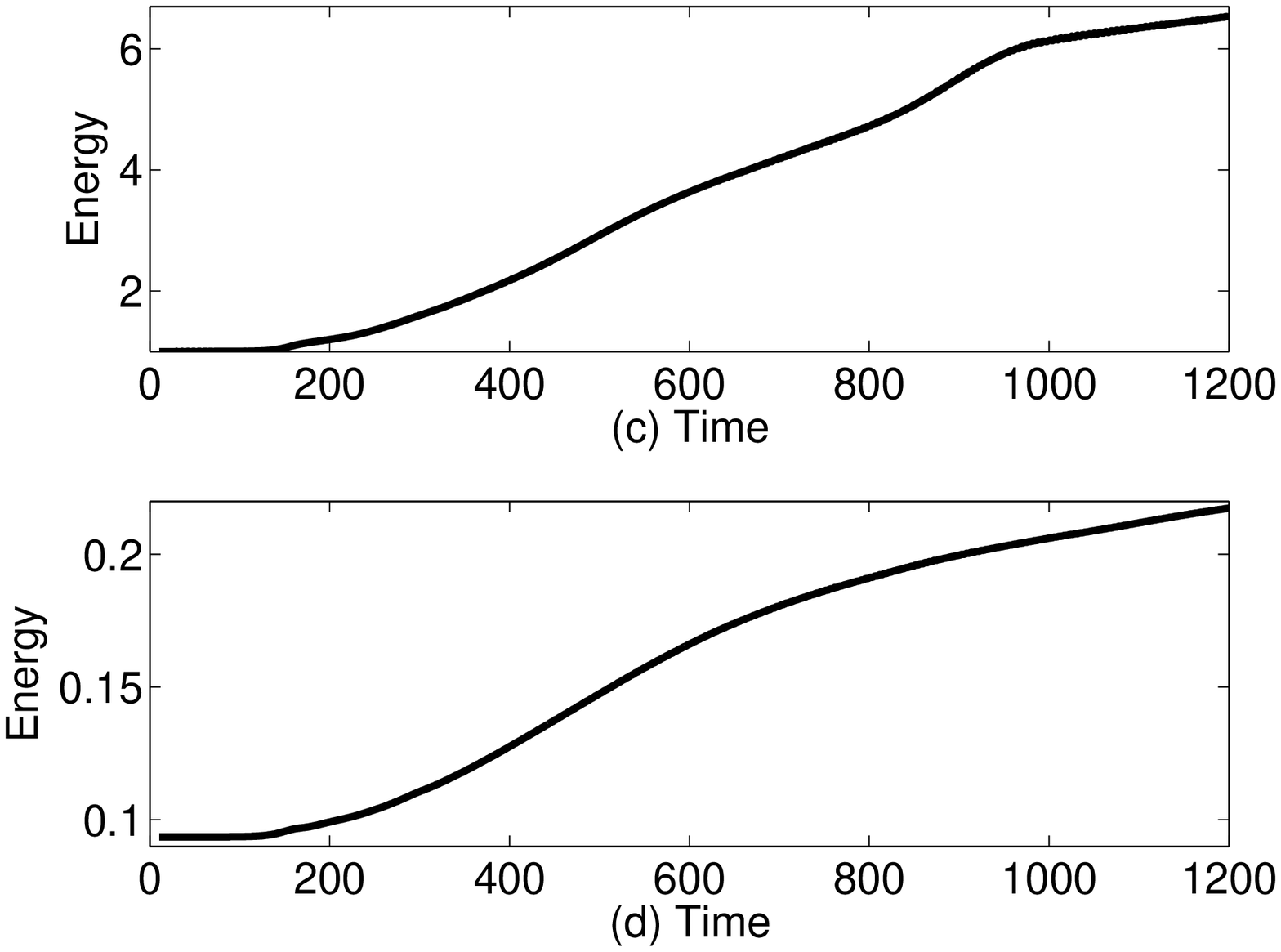}
\caption{(Colour online) The electric $D_{ex}$ (upper curve), $D_{ey}$ and the magnetic $D_{bz}$ 
(lower curve) energies are displayed in panel (a). The dashed blue lines are exponential fits
with the same growth rate. Panel (b) shows the box-averaged electric field amplitude $\langle E_x 
\rangle $. The electron energy $K_e (t)$ is shown in panel (c) and that of the bulk ions in panel 
(d). All energies are normalized to $E_{K0}$.}\label{Fig2}
\end{figure*}
Figure \ref{Fig2} shows the time evolution of their energies. An exponential growth of the electric 
field energy is observed in the time interval $50< t< 150$ with $D_{ex} \approx 10 D_{ey}$. Fitting 
$\exp{(2\omega_i t)}$ to the energy densities $D_{ex}$ and $D_{ey}$ gives us the amplitude's growth 
rate $\omega_i \approx 0.035$. This growth rate is well below the peak $\omega_{i1} \approx 0.058$ 
in Fig. \ref{Fig1}(b). $D_{ex}$ and $D_{ey}$ are integrated over the entire simulation box, which 
averages the field energies over all unstable modes and not just over the fastest-growing ones. The 
saturation of $D_{ex}$ and $D_{ey}$ coincides with the growth of $K_e$ and $K_i$, which evidences the 
saturation of the BTI driven by these two species.  

Figure \ref{Fig2}(b) reveals that $d_t \langle E_x \rangle < 0$ during $100 < t < 150$, which implies 
by Amp\`ere's law a box-averaged ($k_x=0$) net current $j_x > 0$ and thus an electron deceleration in
the reference frame of the simulation box. The net current is tied to a change of the mean speed of 
the electrons in response to the BTI between the bulk ions and electrons and their re-distribution in 
phase space by its saturation. This current is not compensated by a change in the ion mean speed 
due to the large ion inertia. The electrons are accelerated by the $\langle E_x \rangle < 0$, so 
that the electron mean speed along $x$ remains close to $v_e$. The required acceleration of electrons 
by $\langle E_x \rangle$ is small. Even the peak electric field $\langle E_x \rangle \approx -4 
\times 10^{-5}$ can change the electron speed by only $\Delta_v \approx 3 \times 10^{-3}$ during 
a time interval $\Delta_T \approx 1$. A rapid switch to $d_t \langle E_x \rangle > 0$ takes place 
at $t \approx 150$, which coincides with the onset of the growth of the electron and bulk ion 
energies and is thus related to the nonlinear saturation of the BTI between these species. 

All displayed field components and the electron energy continue to grow after $t\approx 150$. $D_{ex}$ 
and $D_{ey}$ grow by another order of magnitude until $t\approx 500$ when they reach their maximum. We 
show below that this growth is caused by the BTI between the fast ion beam and the electrons. The 
growth rates of $D_{ex}$ and $D_{ey}$ are well below that of the initial growth phase prior to 
$t\approx 150$ and $D_{ex}$ is not proportional to $D_{ey}$. We observe a $d_t \langle E_x \rangle >0$ 
and thus a change of the sign of the net current to positive values after $t\approx 350$. Electrons 
start to interact with the waves driven by the second BTI and are accelerated to positive $v_x$. An 
exponential growth of $D_{bz}$ between $200 < t < 400$ is observed and it remains approximately constant 
after this time. The magnetic energy remains 2-3 orders of magnitude below that of the electric field 
components, which implies an essentially electrostatic plasma dynamics. The electron energy $K_e$, 
which takes into account their bulk flow, grows practically linearly up to a value $6 E_{K0}$ in the 
interval between $150 < t < 1000$.

\subsection{Saturation of the instability between electrons and bulk ions}

\begin{figure*}[t]
\includegraphics[width=0.49\columnwidth]{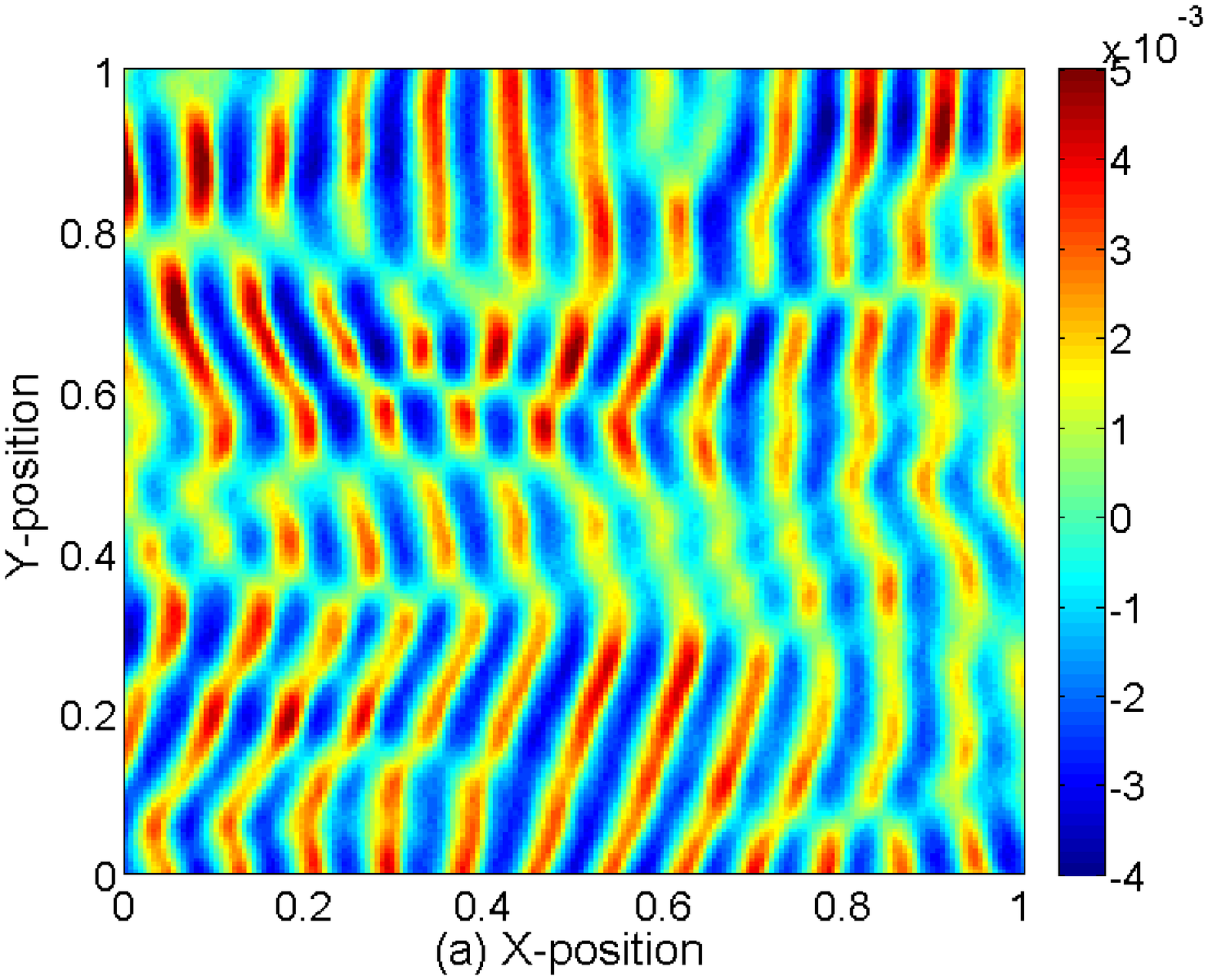}
\includegraphics[width=0.49\columnwidth]{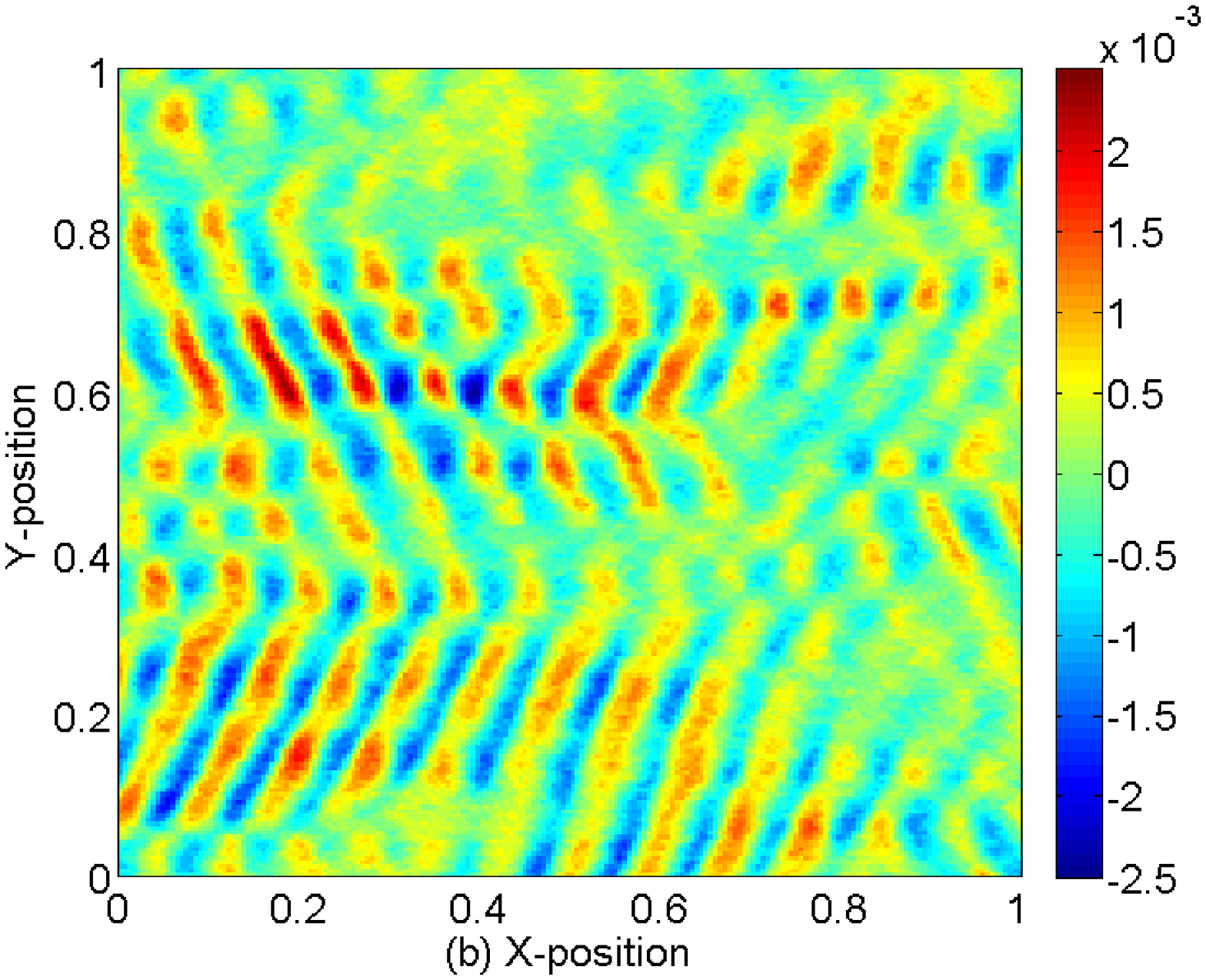}
\caption{(Color online) The electric field distributions in a subinterval of the simulation
box at $t=150$: $E_x$ is shown in panel (a) and $E_y$ in panel (b).}
\label{Fig3}
\end{figure*}

Figure \ref{Fig3} shows the electric fields in a part of the simulation box at 
$t=150$. 
The structures in $E_x$ and $E_y$ belong to waves with wave vectors that are in some cases tilted 
with respect to the beam velocity vector. Their amplitude is two orders of magnitude larger than 
the mean electric field $\langle E_x \rangle$ in Fig. \ref{Fig2}(b). Such wave fields have also 
been observed previously in electrostatic PIC simulations \cite{Amano}. The $E_x$ and $E_y$ are 
projections of the obliquely oriented electric field onto the $x$ and $y$ axes, explaining why 
$D_{ex} \propto D_{ey}$ until $t=150$. The wavelength of the structures is $\lambda \approx 0.06$ 
or $k = {(k_x^2+k_y^2)}^{1/2}\approx 100$. This value is in line with the wave numbers of the 
oblique modes in Fig. \ref{Fig1}(b).

\begin{figure*}[t]
\includegraphics[width=0.49\columnwidth]{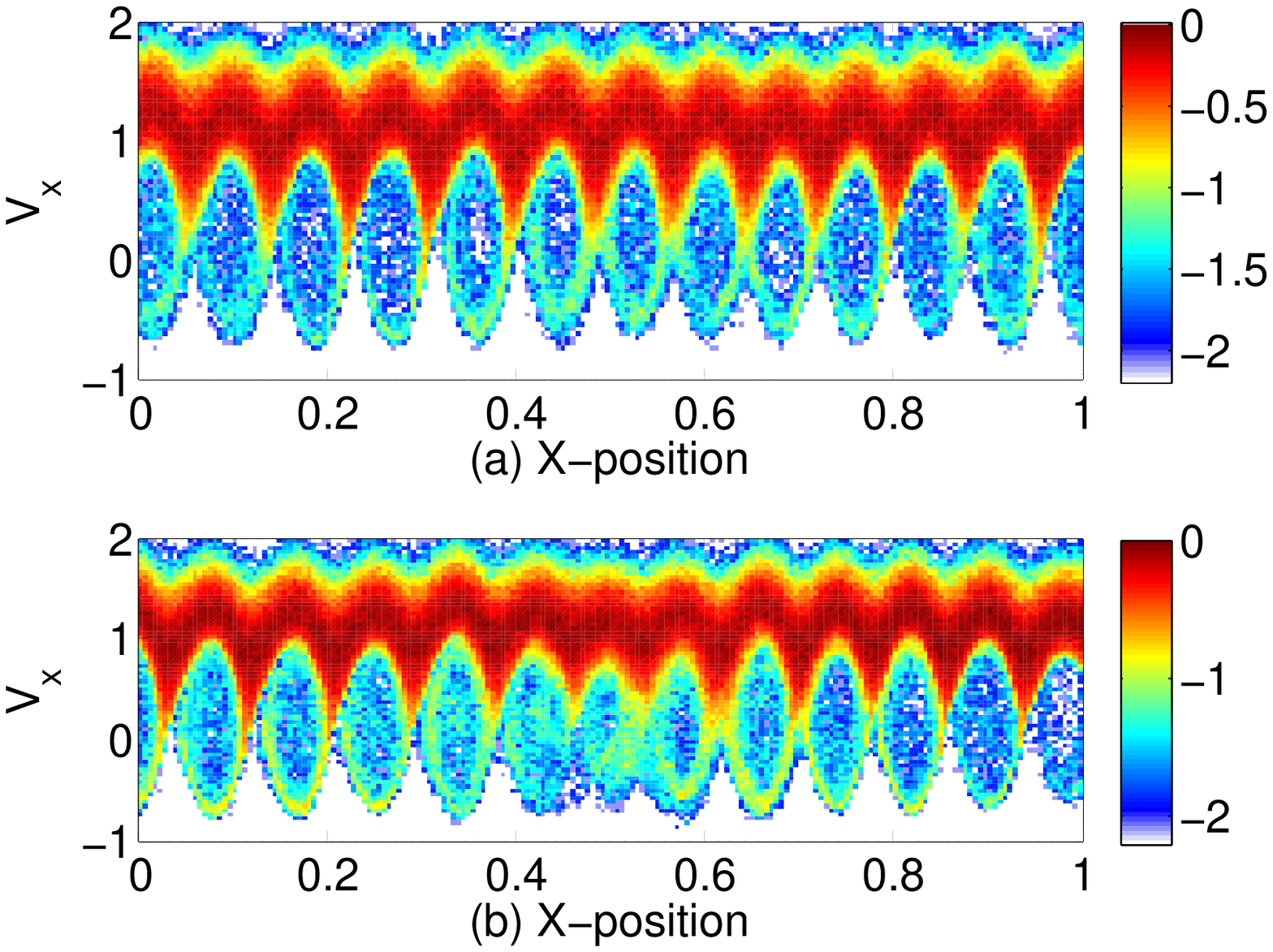}
\includegraphics[width=0.49\columnwidth]{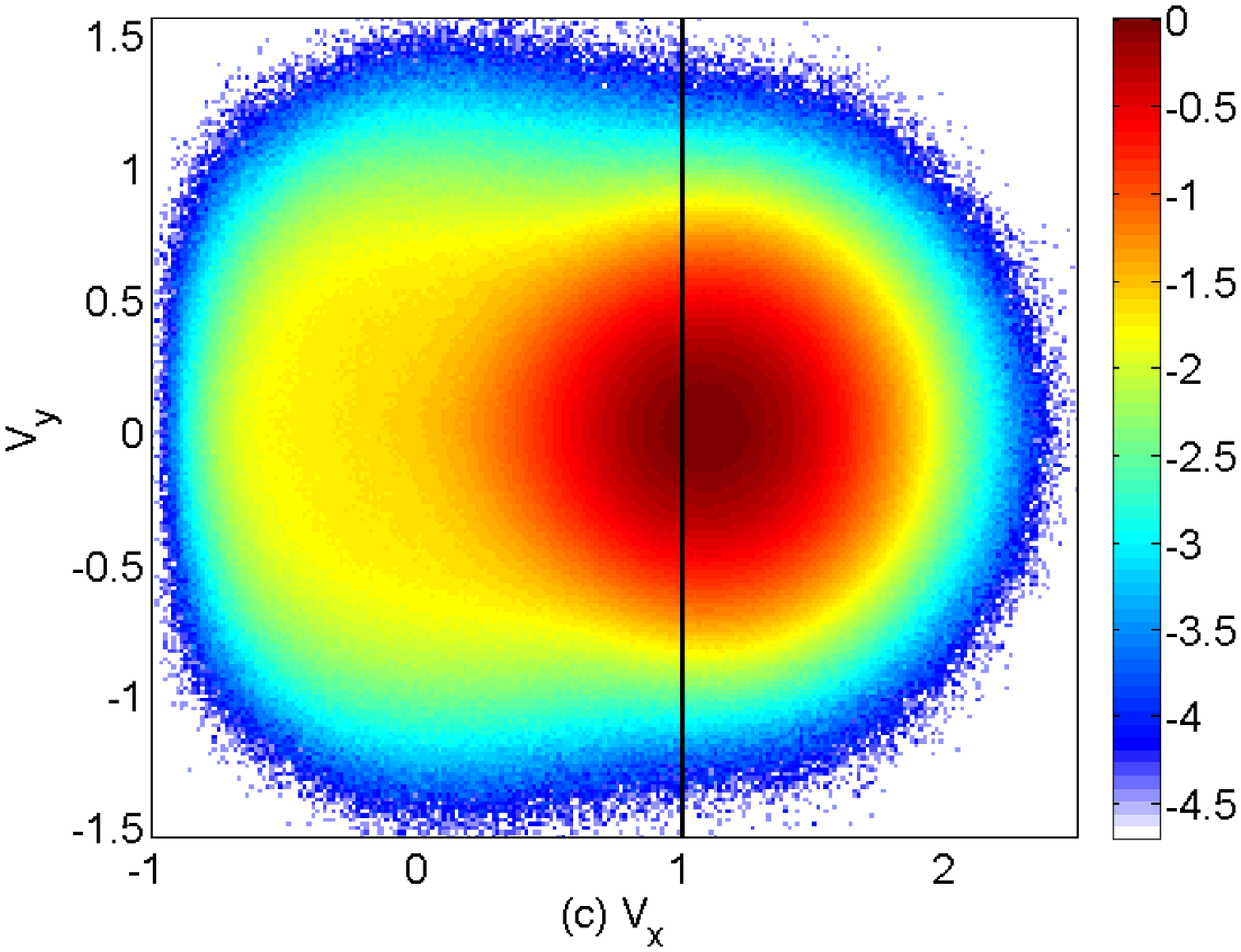}
\caption{(Color online) Electron phase space distributions at $t=150$: Panel (a) and (b)
show phase space projections onto the $x,v_x$ plane. The distribution is integrated over 
$0 < y < 0.05$ in (a) and over $1.55 < y < 1.6$ in (b). The electron distribution has been 
integrated over all $x,y,v_z$ for the projection onto the $v_x,v_y$ plane in panel (c). 
The vertical line corresponds to $v_x = v_e$. All distributions are normalized to their 
respective maxima and the colour scale is 10-logarithmic.}
\label{Fig4}
\end{figure*}

Various projections of the resolved electron phase space distribution $f(x,y,v_x,v_y,v_z)$ at 
$t=150$ are shown in Fig. \ref{Fig4}. The phase space projection on the $x,v_x$ plane integrates 
this distribution over all $v_y,v_z$ and over ten grid cells along $y$. Figures \ref{Fig4}(a) and 
(b) integrate the distribution over $0 < y < 0.05$ and over $1.55 < y < 1.6$, respectively.  
Figures \ref{Fig4}(a) and (b) reveal electron phase space holes, which arise when electrostatic 
waves saturate \cite{H1}. The spatial width of each phase space hole corresponds to one wave 
period of the electrostatic waves driven by the BTI. The exponential growth phase of $D_{ex}$ 
and $D_{ey}$ thus ends at $t=150$ as usual due to the trapping of electrons. The electrons are 
trapped by a potential, 
which is practically stationary in the rest frame of the bulk ions. This is evidenced by the
supplementary movie 1, which shows a time animation of Fig. \ref{Fig4}(a). The electrons gyrate
around $v_x = 0$ at early times. The mean speed of the trapped electrons inside of a phase space 
hole is less than $v_e$, which can be seen in Fig. \ref{Fig4}. A significant fraction of the 
electrons has been slowed down at $t=150$, which reduces the modulus of the electron current. 
The electron current is negative because $v_e > 0$. The ions hardly react to the electric field 
because $m_i \gg m_e$ and they can not compensate the change of the electron current. We obtain 
a macroscopic box-averaged $\langle j_x \rangle > 0$, which explains why $d_t \langle E_x \rangle 
< 0$ in Fig. \ref{Fig2}(b). The mean electric field $\langle E_x \rangle < 0$ in Fig. \ref{Fig2}(b) 
accelerates all electrons in the positive direction, which will reduce $\langle j_x \rangle$. The 
maximum of the phase space density in Fig. \ref{Fig4}(c) is thus found at $v_x > 1$. The electrons 
in this core population have not been trapped as we can see from Figs \ref{Fig4}(a,b) and from 
movie 1, but they have all gained speed along $x$. The trapped electrons in Fig. \ref{Fig4}(c) 
are found at $v_x < 1$. The electron distribution is stretched out along $v_y$, because the 
electrons are trapped by oblique waves (See Fig. \ref{Fig3}).

\begin{figure*}[t]
\includegraphics[width=0.49\columnwidth]{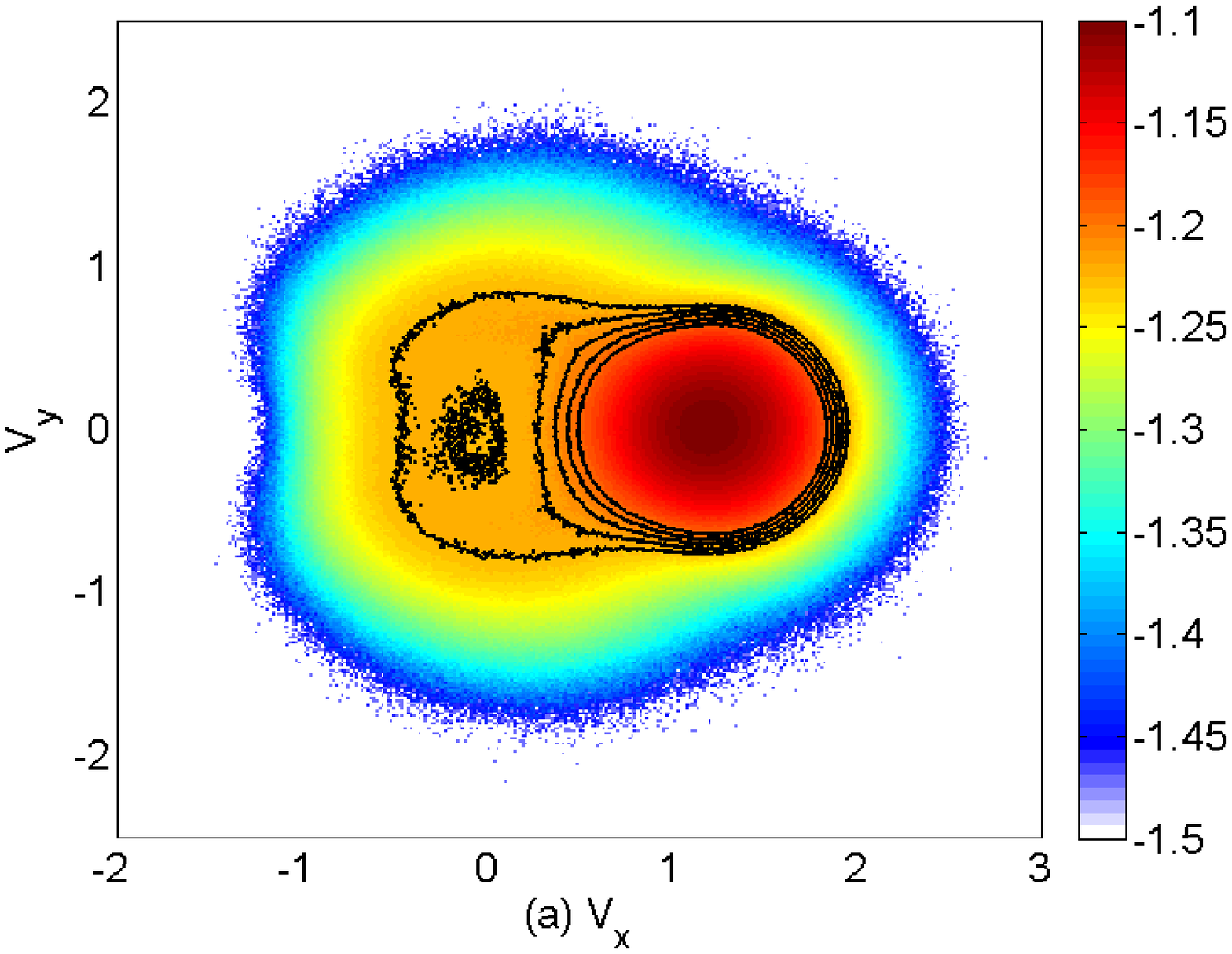}
\includegraphics[width=0.49\columnwidth]{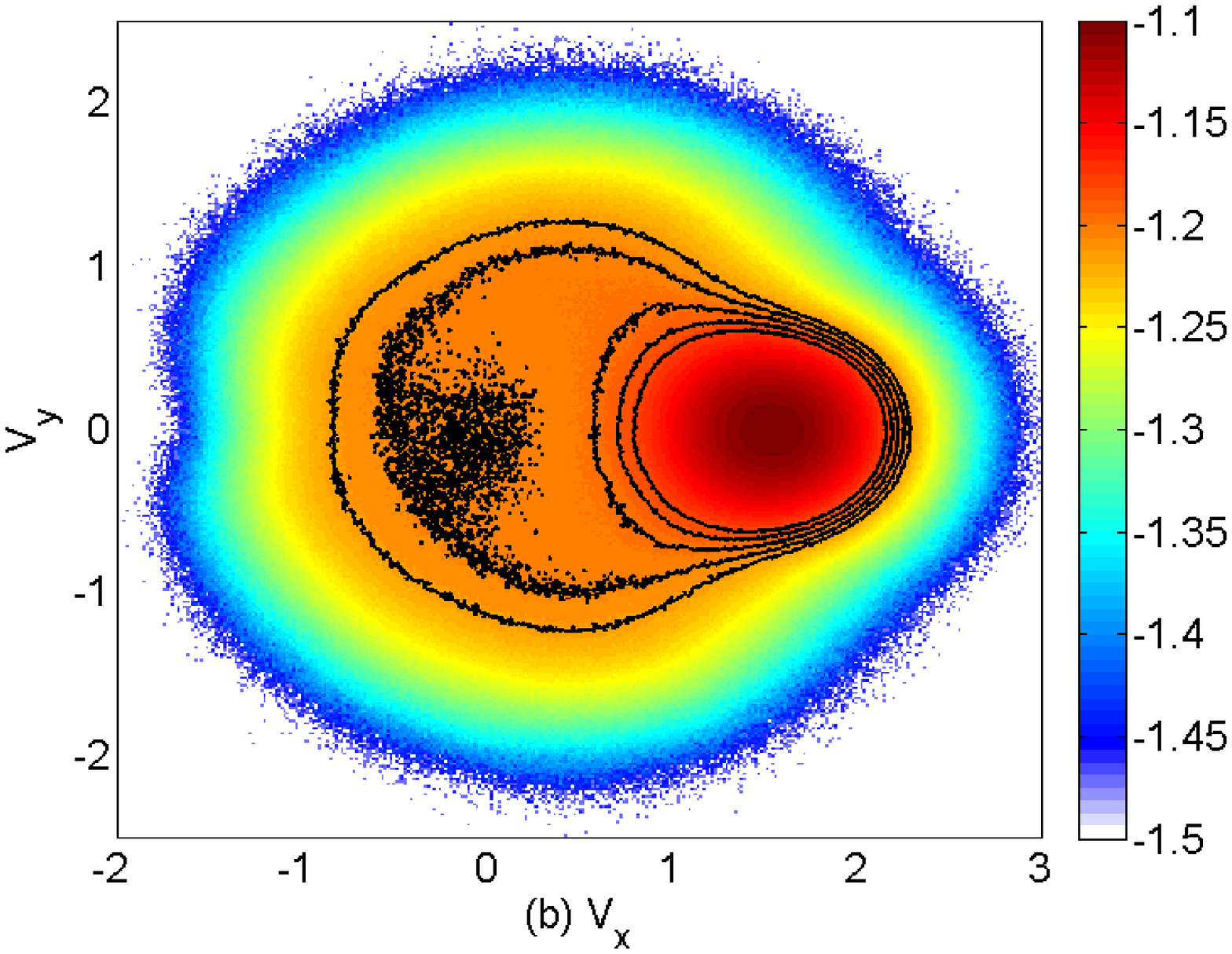}
\caption{(Color online) Evolution of the electron velocity distribution: Panel (a) shows 
the electron distribution at $t=200$ and panel (b) shows it at $t=300$. Both distributions
are normalized to the peak value at $t=150$ and the colour scale is 10-logarithmic. The 
black curves are the contour lines -1.24, -1.23, -1.22, -1.21, -1.20.}\label{Fig5}
\end{figure*}

The electrons are accelerated by $\langle E_x \rangle < 0$ (Fig. \ref{Fig2}(b)) into the 
positive x-direction and they are heated by their interaction with the localized electric 
field structures (Fig. \ref{Fig3}). Figure \ref{Fig5} demonstrates this by a comparison 
of the electron velocity distributions at the times $t=200$ and $t=300$, when the modulus 
of $\langle E_x \rangle$ is largest.
The velocity that corresponds to the maximum of the electron phase space density is found 
at $v_x \approx 1.2$ at $t=200$ and at $v_x \approx 1.6$ at $t=300$. The distribution widens in velocity space, which corresponds to an increasing thermal
speed, and it adopts an increasingly circular shape. The centre of the hot circular distribution 
is at $v_x \approx 0.3$ at $t=300$. We can not observe a density distribution in Fig. \ref{Fig5} 
that resembles the crescent found in Ref. \cite{Amano}.

The electron distribution in Fig. \ref{Fig4}(c) can be subdivided into a thermal core 
population centred at $v_x \approx 1.2$ and $v_y \approx 0$ with a high density and a 
thermally anisotropic hot electron population with a lower density. The low-density electron 
population is elongated along $v_x$ in Fig. \ref{Fig5}(a) and it has become almost circular at 
$t=300$ in Fig. \ref{Fig5}(b). It is thus fair to assume that this thermally anisotropic electron 
population is responsible for the growth of $D_{bz}$ in Fig. \ref{Fig2}(a) and that the magnetic 
field growth ceases once the electron distribution is isotropic in velocity space. A thermally 
anisotropic electron distribution drives waves with a wave vector, which is aligned with the 
direction along which the electrons are cool \cite{Weibel}. We thus expect a wave vector of the 
magnetowaves that is almost parallel to the y-direction. The magnetic instability is driven by 
currents in the x-y plane and the unstable magnetic field component is thus aligned with the 
z-direction. Indeed only $D_{bz}$ grows in our simulation, while the other magnetic field 
components remain at noise levels (not shown). 

\begin{figure*}[t]
\includegraphics[width=0.32\columnwidth]{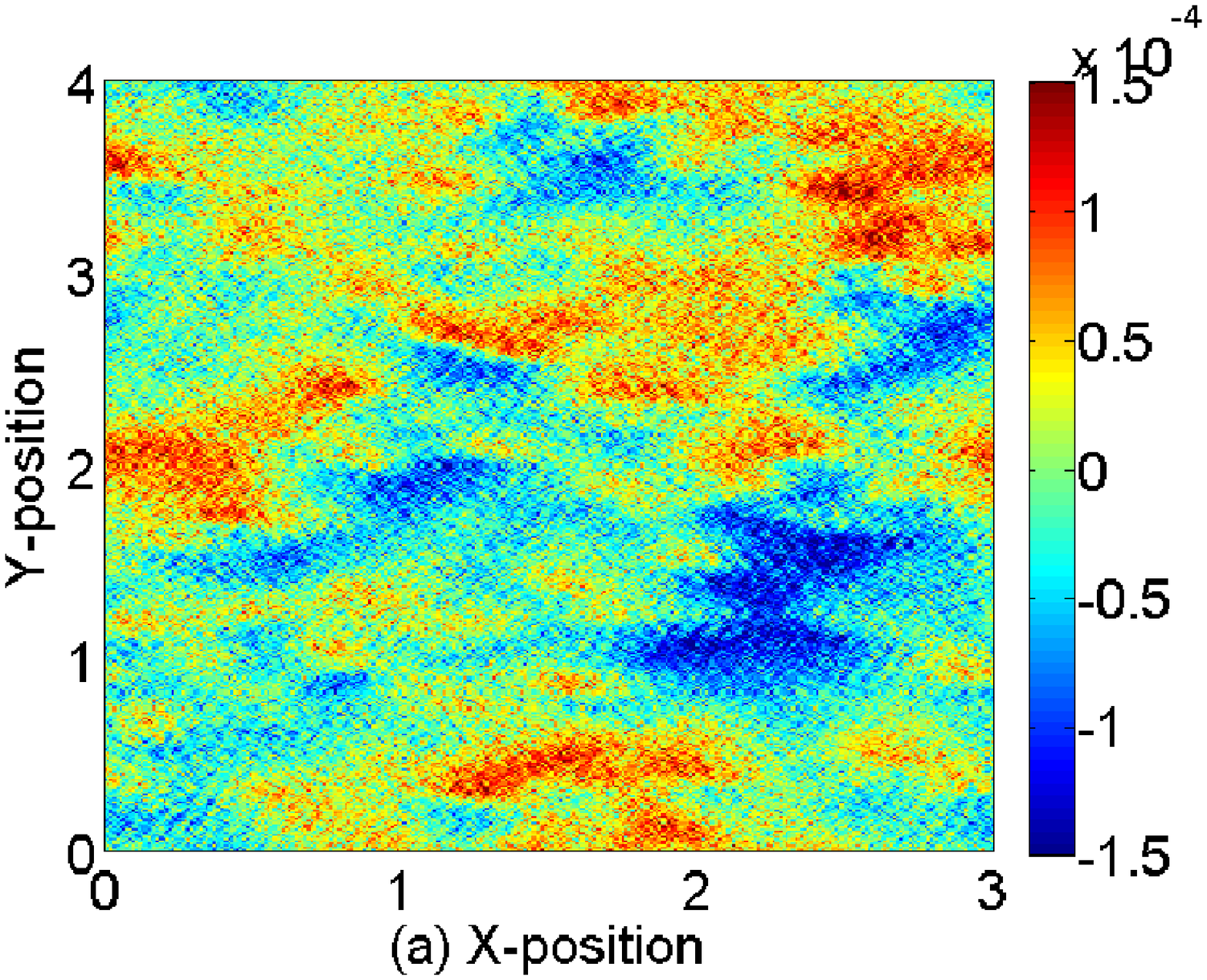}
\includegraphics[width=0.32\columnwidth]{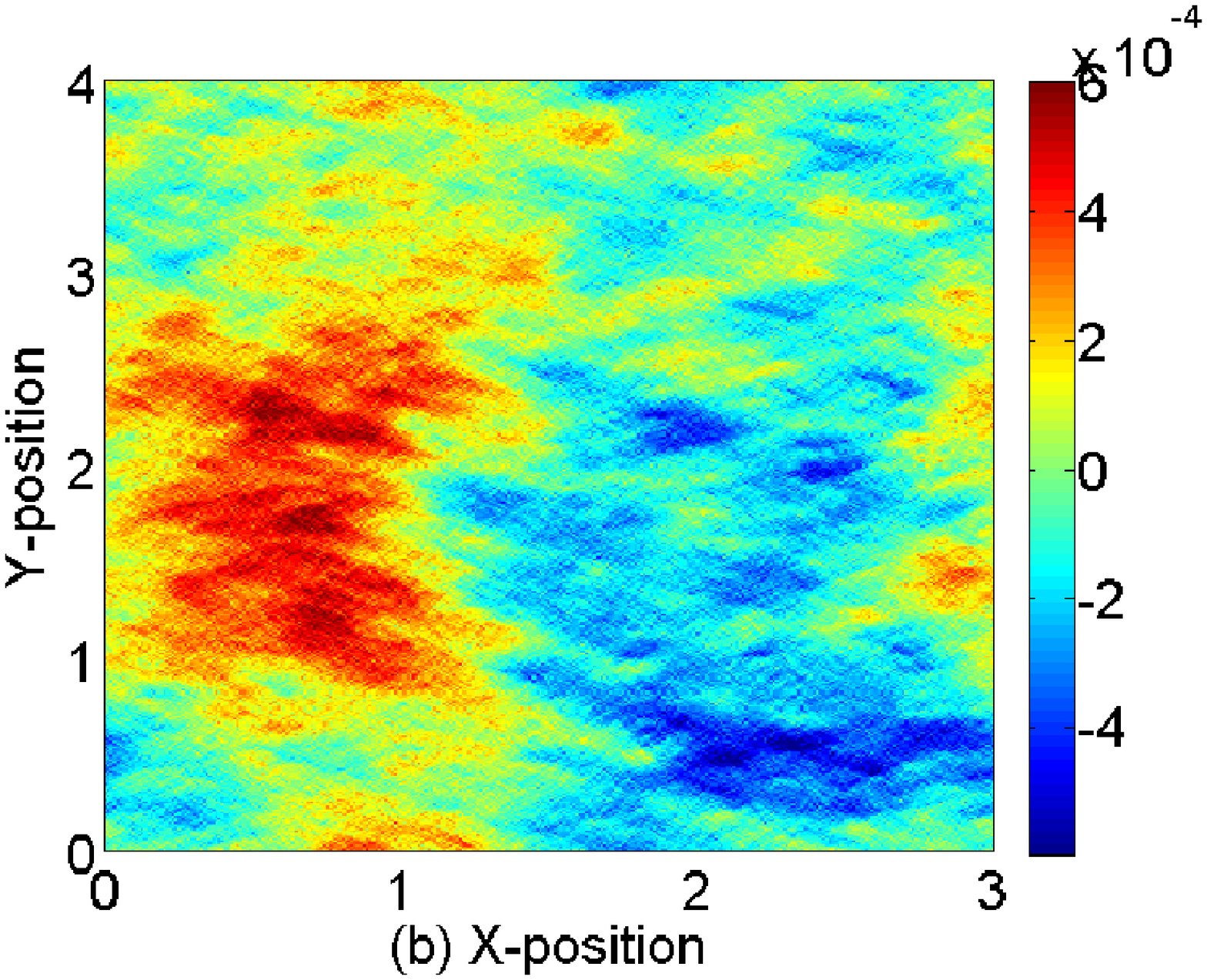}
\includegraphics[width=0.32\columnwidth]{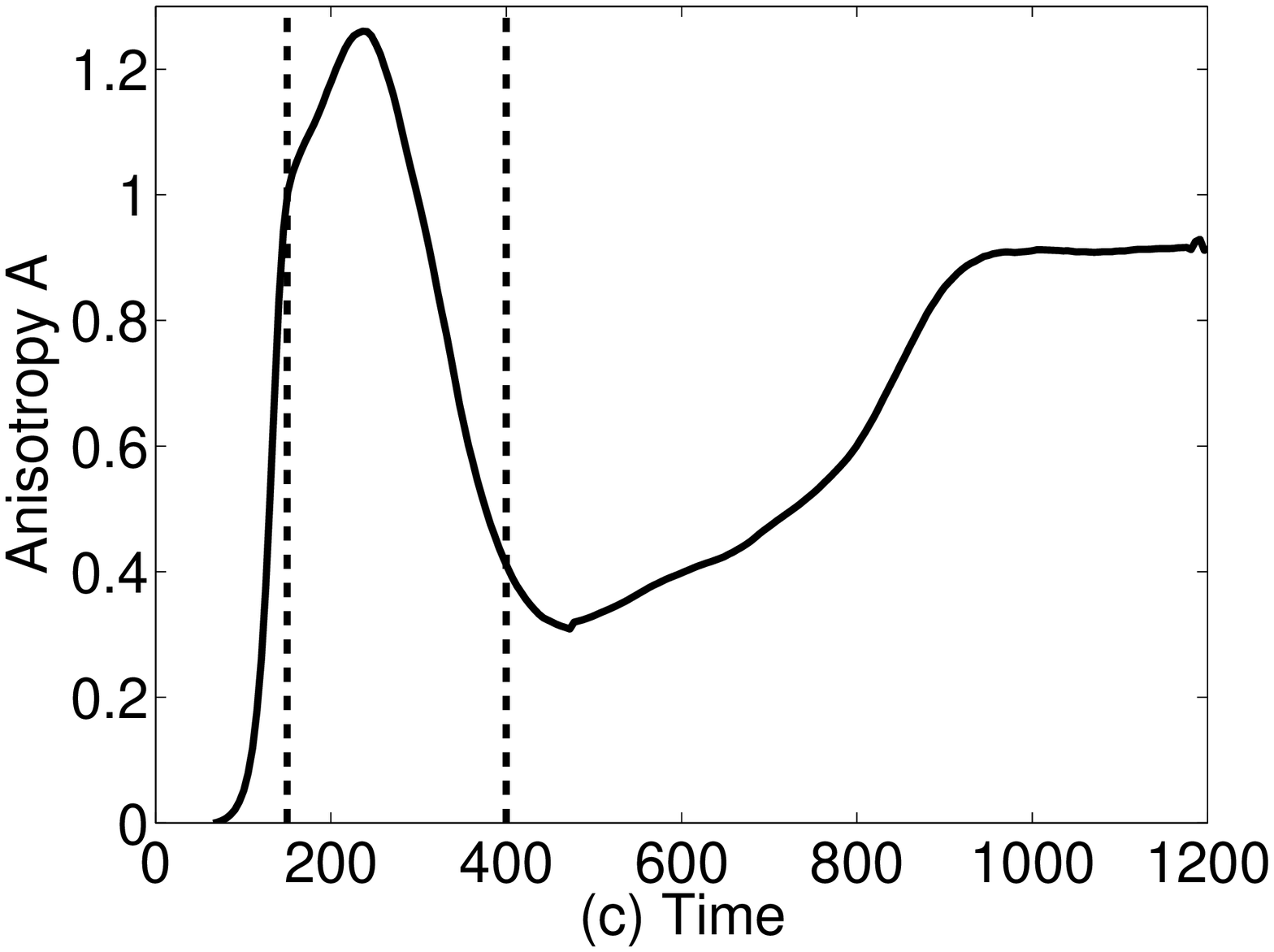}
\caption{(Color online) The $B_z$ component of the magnetic field at the times $t=150$ (a) and 
$t=400$ (b). The full simulation grid is displayed. The thermal anisotropy $A$ is shown as a 
function of time $\omega_p t$ in (c) and the times $t=150$ and $t=400$ are overplotted as
vertical dashed lines.}\label{Fig6}
\end{figure*}

Figure \ref{Fig6} shows the electron thermal anisotropy $A(t)$ and compares the spatial distribution of $B_z$ at the time $t=150$, which is just 
before the exponential growth phase of $D_{bz}$ in Fig. \ref{Fig2}(a), and at $t=400$, when the 
exponential growth phase of $D_{bz}$ ends.
The magnetic field is initially weak and the structures are small. The saturation of the BTI between 
the bulk plasma and the electrons has resulted in a rapid growth of $A$, which has reached $A \approx 
1$ at $t=150$. This value corresponds to an electron thermal energy along $x$ that is twice that along $y$. 
The anisotropy reaches its peak at $t \approx 250$ and decreases thereafter to a value $A\approx 0.4$
at $t=400$, when a large coherent magnetic field structure has emerged in Fig. \ref{Fig6}(b). The size 
of this patch is of the order of an electron skin depth in both spatial directions and the magnetic 
amplitude gives a ratio between the electron cyclotron frequency and the plasma frequency of $6 \times 
10^{-4}$, which is comparable to the ratio found in the interstellar medium \cite{Volk}. The field 
patch is not thermal noise. Thermal noise \cite{PhysScripta} is linked to charge and current density 
fluctuations due to the finite number of computational particles and its coherence length is comparable 
to the grid cell size. The structure centred at $x=0.7$ and $y=1.7$ in \ref{Fig6}(b) is, however, not 
the plane wave with a wave vector parallel to the y-axis, which we would expect from a TAWI 
\cite{Stockem1}. A magnetic structure is growing while $A$ is large, but we can not attribute it to an elementary 
instability such as a TAWI.

\subsection{Saturation of the instability between electrons and the fast ion beam}

The comparison of Figs. \ref{Fig1}(a) and \ref{Fig5}(b) demonstrates that the mean electric field 
$\langle E_x \rangle$ in Fig. \ref{Fig2}(b) has accelerated the bulk of the electrons from $v_x = 
1$ to $v_x \approx 1.7$. The velocity gap $\Delta v_x$ between the value $v_x$, where the electron 
phase space density reaches its maximum, and the mean speed $v_b = 6$ of the fast beam has thus 
decreased from 5 to 4.3. The change of $\Delta v_x$ should result in a larger wave number of the 
unstable electrostatic waves than the now normalized $k_{u2} = 1/(v_b - v_e) \approx 13$ estimated 
in section 2. This relative speed $\Delta v_x$ expressed in units of the electron thermal speeds 
is also well below the value used in Ref. \cite{Amano}, in particular if we take into account 
that the electrons have been heated up by the saturation of the first BTI. 

\begin{figure*}[t]
\includegraphics[width=0.49\columnwidth]{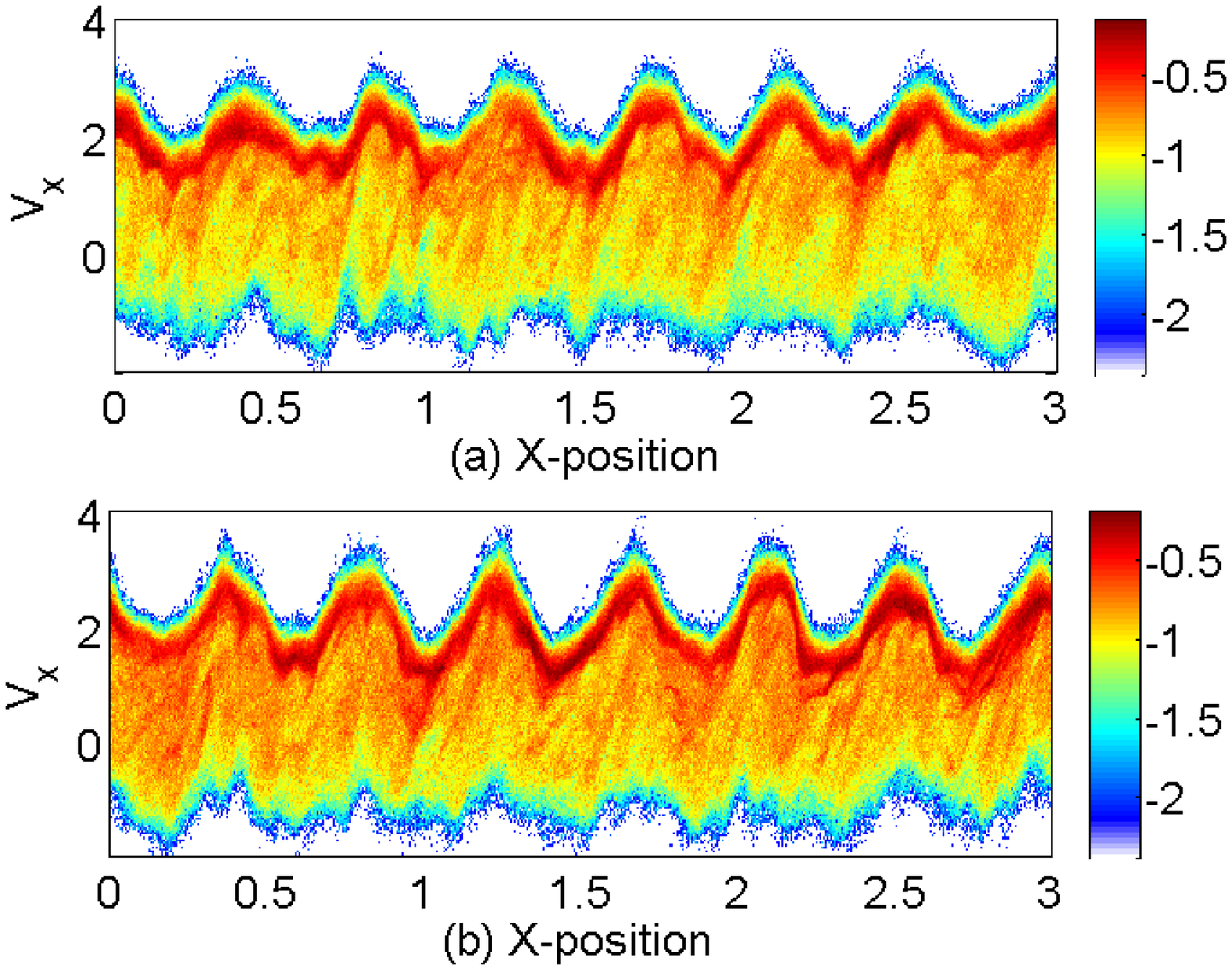}
\includegraphics[width=0.49\columnwidth]{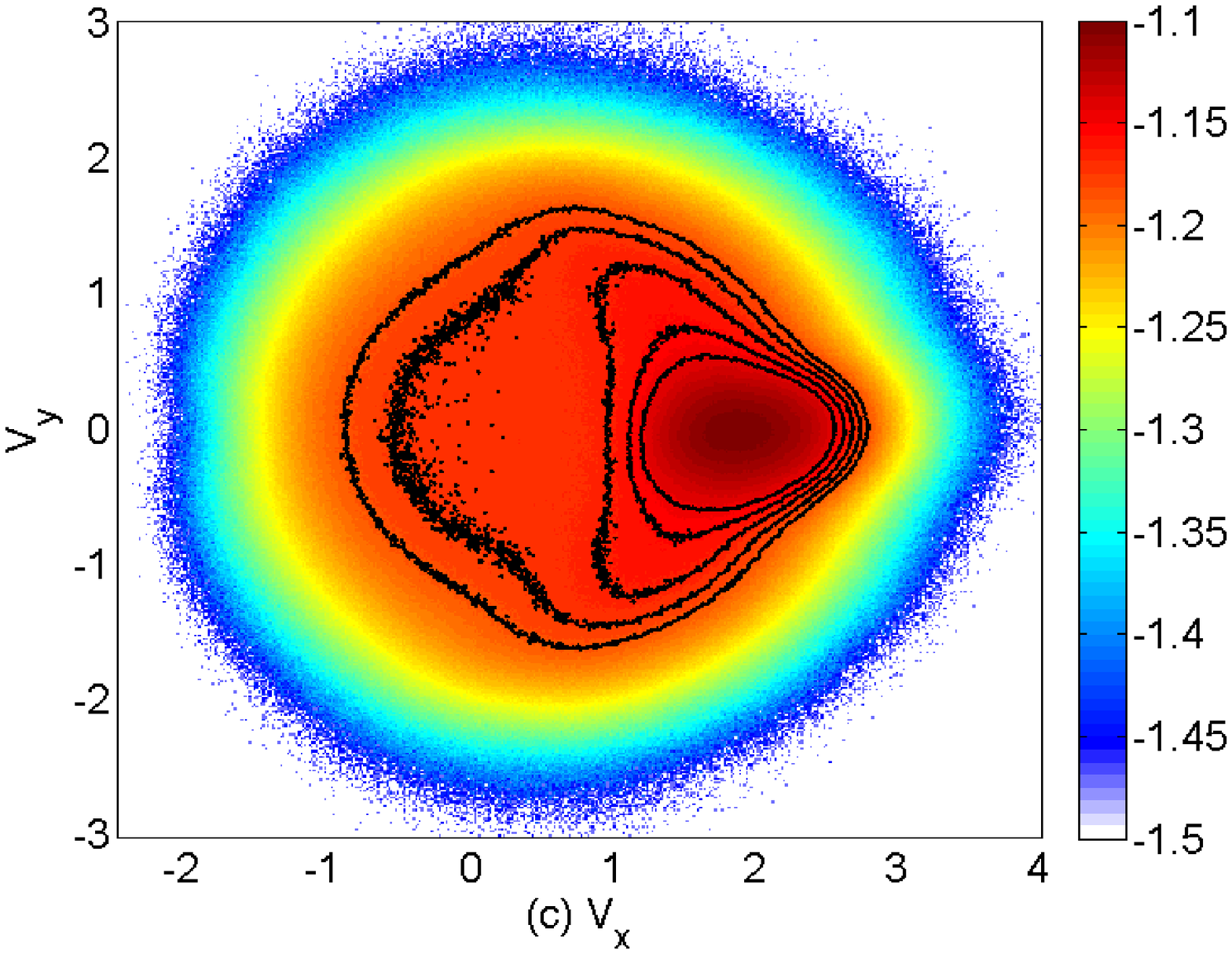}
\caption{(Color online) Electron phase space distributions at $t=400$: Panel (a) and (b) show phase 
space projections onto the $x,v_x$ plane. The distribution is integrated over $0 < y < 0.05$ in (a) 
and over $1.55 < y < 1.6$ in (b). The electron distribution has been integrated over all $x,y,v_z$ 
for the projection onto the $v_x,v_y$ plane in panel (c). All distributions are normalized to their 
respective maxima at $t=150$ and the colour scale is 10-logarithmic. The black curves in (c) are the 
contour lines -1.2, -1.19, -1.18, -1.17, -1.16.}\label{Fig7}
\end{figure*}

Figures \ref{Fig7}(a,b) reveal that this second BTI has already commenced to grow at $t=400$. 
Movie 1 shows a wave perturbation after $t\approx 300$ of the electron distribution that 
propagates at a high speed to increasing $x$, which corresponds to the wave driven by the BTI 
between the electrons and the fast beam. Strong modulations of the electron distribution with 
$1.5 < v_x < 3$ are present. Their wave number is $k_o = 14 \pi / 3 \approx 14.6$, which is 
indeed larger than $k_{u2}$. However, $k_o$ may not correspond to the true fastest growing wave 
because the periodic boundary conditions enforce a discrete wave spectrum.   
The electron velocity distribution in Fig. \ref{Fig7}(c) reveals that the dilute electron component 
forms an almost circular distribution centred at $v_x \approx 0.5$ and $v_y\approx 0$. The contour 
line corresponding to $10^{-1.18}$ at $v_x \approx 1$ is almost aligned with the $v_y$-axis. It 
resembles the crescent in Ref. \cite{Amano}, but it is much less pronounced.  

\begin{figure*}[t]
\includegraphics[width=0.49\columnwidth]{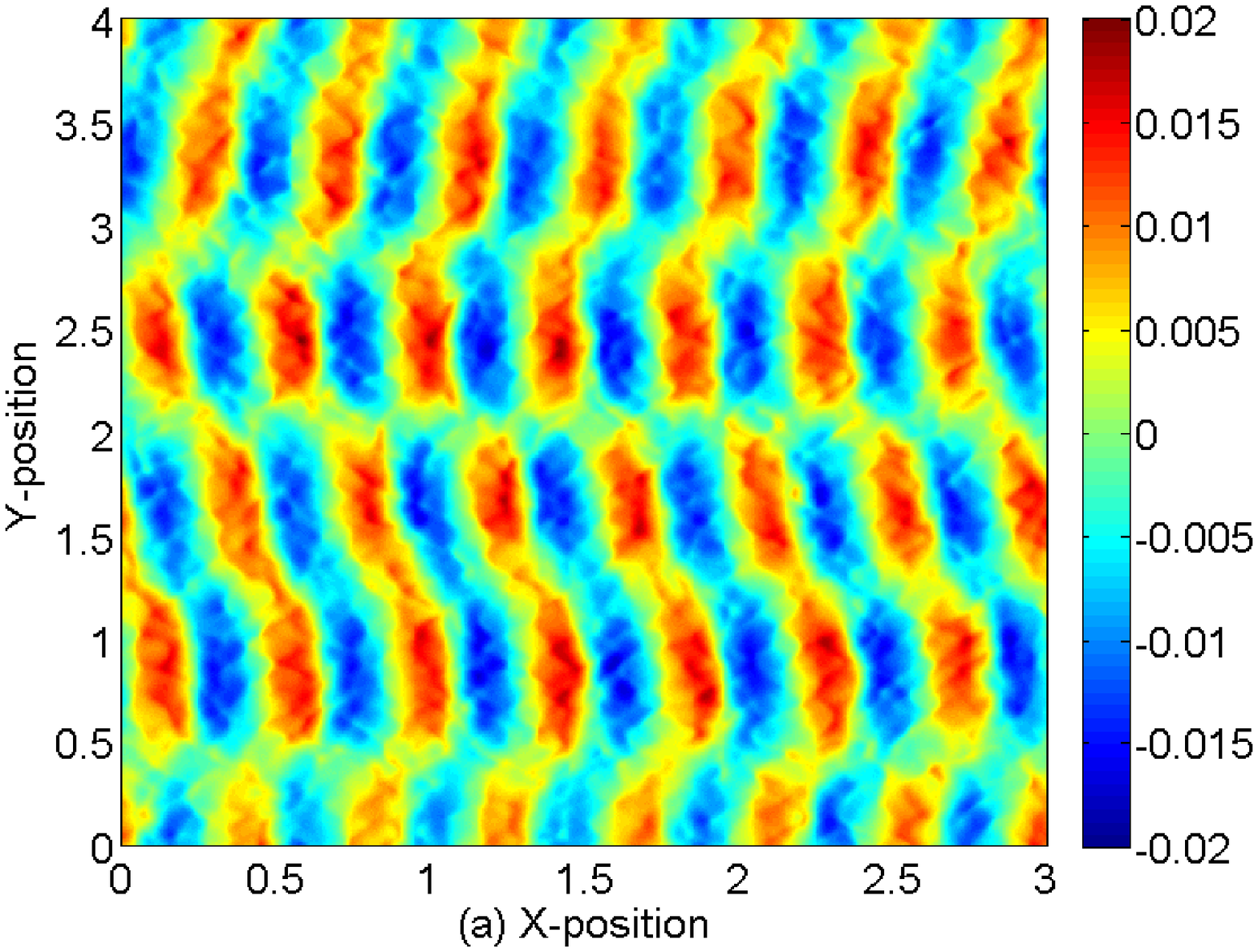}
\includegraphics[width=0.49\columnwidth]{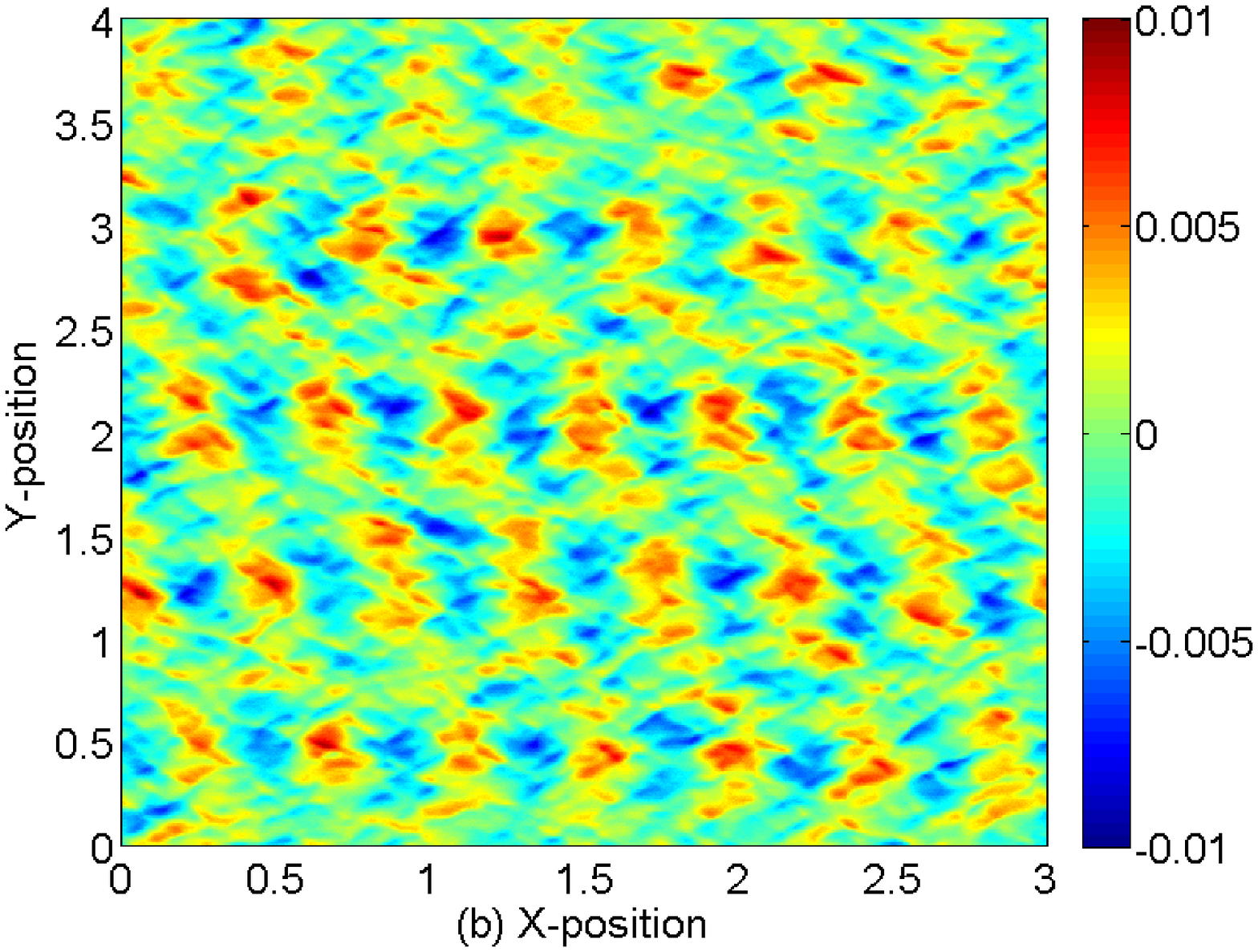}
\caption{(Color online) The electric field distributions at $t=500$: $E_x$ is shown in 
panel (a) and $E_y$ in panel (b).}\label{Fig8}
\end{figure*}

Figure \ref{Fig8} shows the electric field distribution at $t=500$, when $D_{ex}$ and 
$D_{ey}$ reach their peak values in Fig. \ref{Fig2}(a).
The structures in $E_x$ and $E_y$ show a spatial correlation due to the obliquity of the wave 
front. The characteristic tilt angle between the wave vector and the x-axis is less than the 
one in Fig. \ref{Fig2} and thus $|E_y| \ll |E_x|$. Internal filamentary structures are visible 
in the intervals with $|E_x| \gg 0$. The filamentary structures are relatively stronger in 
$E_y$ compared to $E_x$, which may explain why $D_{ex}$ does no longer grow in unison with 
$D_{ey}$ in Fig. \ref{Fig2}(a) at this time. The electric field amplitude in Fig. \ref{Fig8}(a) 
exceeds that in Fig. \ref{Fig3}(a) by a factor of 4 and the wave length is larger by a factor 5. 
The electrostatic potential, which is driven by the BTI between the electrons and the fast ion beam, 
thus exceeds that of the faster growing BTI by a factor of $\approx$ 20 and should give trapped 
electron islands larger than those in Fig. \ref{Fig4}. 

\begin{figure*}[t]
\includegraphics[width=0.49\columnwidth]{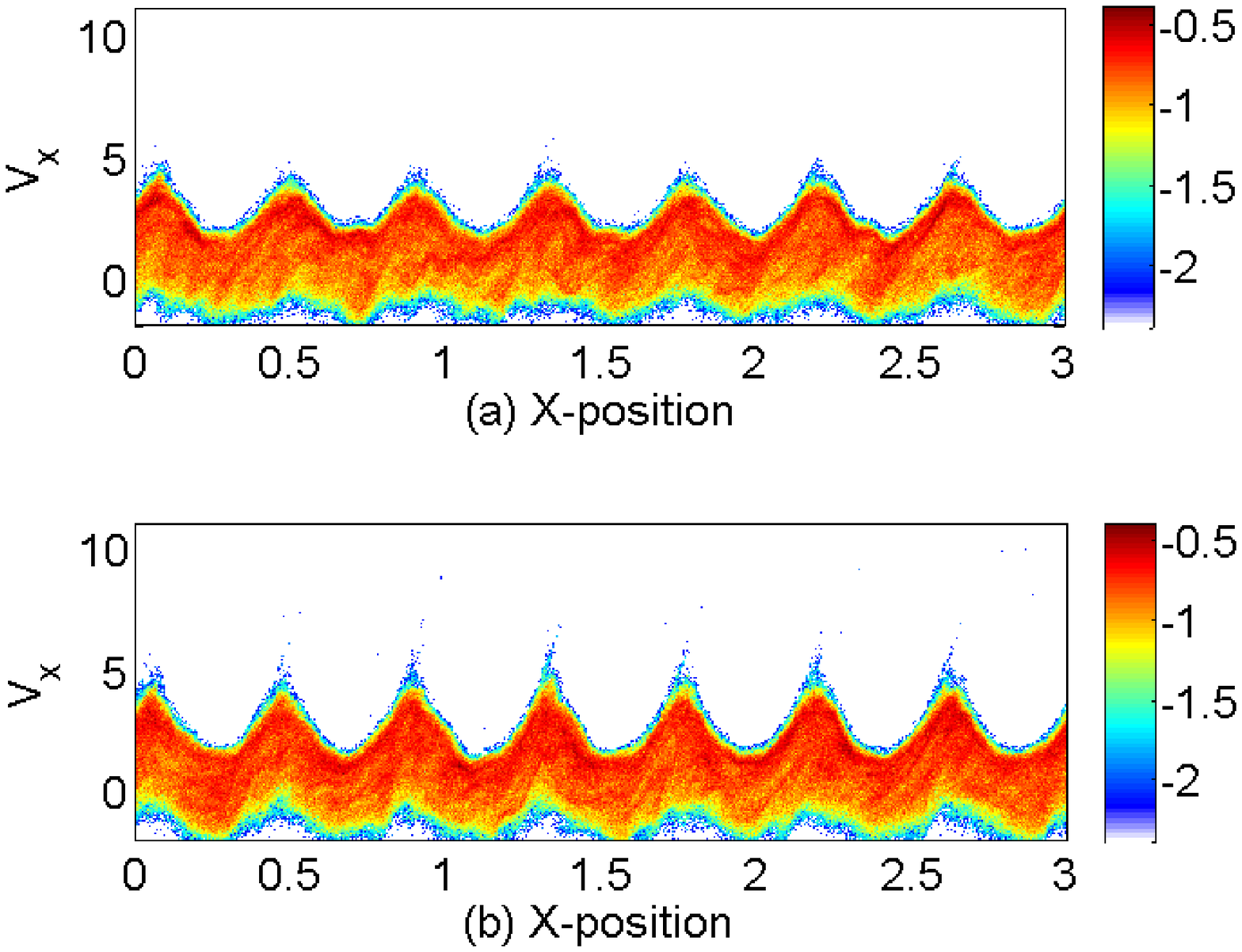}
\includegraphics[width=0.49\columnwidth]{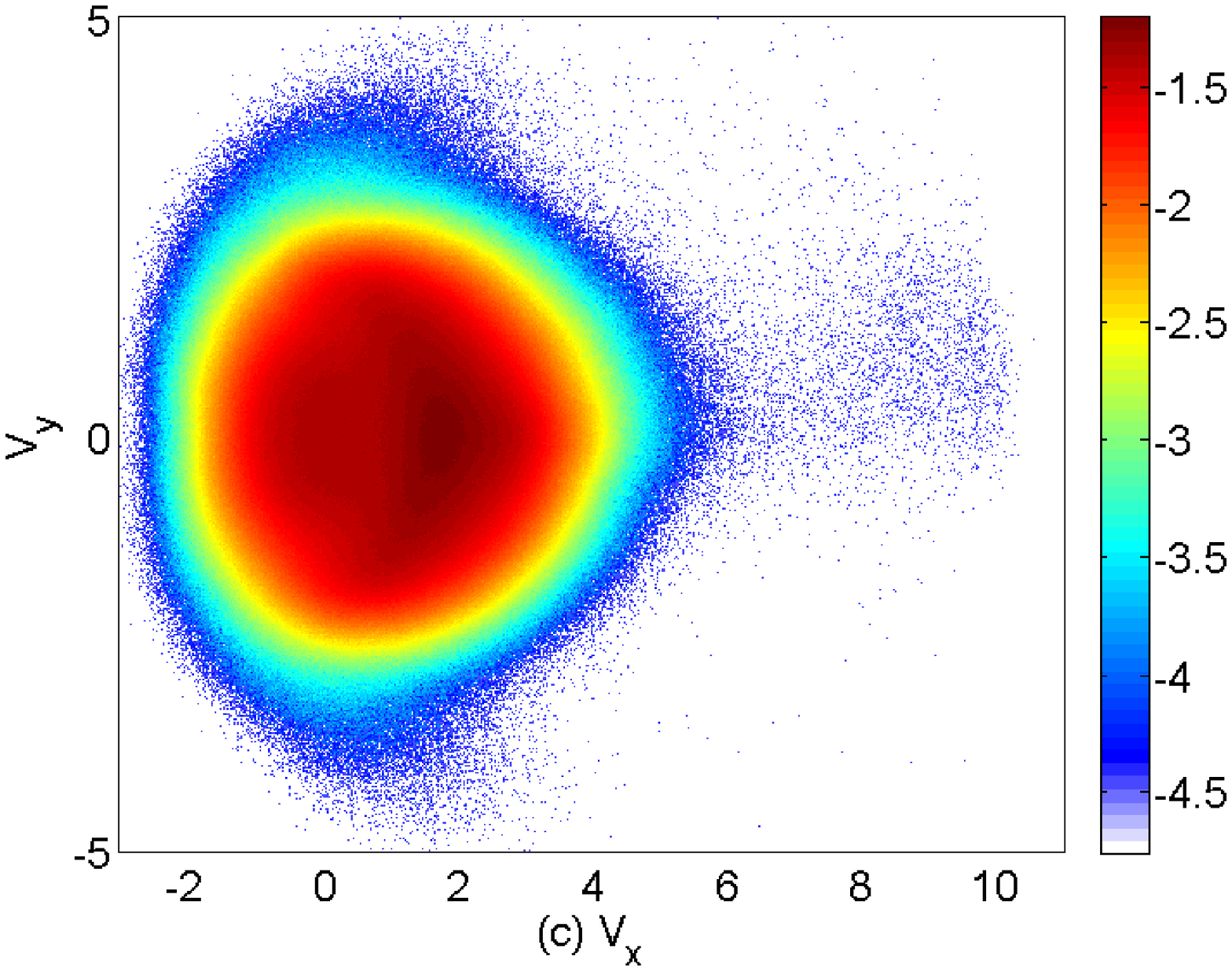}
\caption{(Color online)  Electron phase space distributions at $t=500$: Panel (a) and (b) show phase 
space projections onto the $x,v_x$ plane. The distribution is integrated over $0 < y < 0.05$ in (a) 
and over $1.55 < y < 1.6$ in (b). The electron distribution has been integrated over all $x,y,v_z$ 
for the projection onto the $v_x,v_y$ plane in panel (c). All distributions are normalized to their 
respective maxima at $t=150$ and the colour scale is 10-logarithmic.}
\label{Fig9}
\end{figure*}

Some electrons are trapped at the time $t=500$ by this faster wave, which is the time when 
$D_{ex}$ saturates in Fig. \ref{Fig2}(a). The supplementary movie 2, which shows the spatially 
integrated phase space distribution as a function of $v_x,v_y$ and corresponds to Fig. 
\ref{Fig9}(c), demonstrates that the number of trapped electrons rapidly increases after this time. 
The black lines correspond to the speeds of both ion beams. Figures \ref{Fig9}(a,b) show that 
the electron velocity distribution is not sinusoidal, which is a sign of a nonlinear wave, and that 
some electrons have speeds of $v_x > 5$. These electrons are found close to the cusps of the 
electron distribution with $v_x \approx 5$, which correspond to the unstable equilibrium point of a 
periodic electrostatic potential that moves with a speed $\approx v_b$. The fastest electrons have 
just started their periodic motion in the electrostatic potential of the wave fronts with a wave vector, 
which is parallel to the x-axis. 

The electron velocity distribution in Fig. \ref{Fig9}(c) reveals that a small fraction of electrons 
have been accelerated up to $v_x \approx 10$. These electrons have reached the stable equilibrium 
point of the electrostatic potential and their kinetic energy reaches its peak value. The electron 
distribution extends here and at later times (Movie 2) over a velocity interval that does not 
exceed $20 v_e$ or 0.3c and the electrons thus remain nonrelativistic. The velocity 
distribution reveals non-thermal electrons at $v_x \approx 1$ and at $|v_y| \approx 4$.  
These electrons must have gained speed by their interaction with the oblique modes. They can interact 
more easily with electrons than the beam aligned mode, because their phase speed is lower \cite{Amano}. 
The interaction of electrons with the oblique modes results in a velocity distribution, which resembles 
a crescent. This saturation mechanism is dominant, if the thermal spread of the electrons is small 
compared to the beam speed. The beam speed was 10 times the electron thermal speed in Ref. \cite{Amano}. 
If the electron thermal spread is comparable to the beam speed, like for the instability between the 
bulk ions and the electrons, then the beam aligned modes can easily trap the electrons in the 
high-energy tail of the electron distribution and no crescent develops. The larger gap between the 
speed $v_x \approx 3.5$ of the fastest electrons in Fig. \ref{Fig7}(c) and the ion beam speed $v_b = 
6$ apparently puts the electron interaction with oblique modes and with beam-aligned modes on an equal 
footing. Movie 2 clearly shows that electrons are accelerated 
along $v_x$ at large $v_x$ and along $v_y$ for $v_x \approx 0$ after $t\approx 450$. 

\begin{figure*}[t]
\includegraphics[width=0.49\columnwidth]{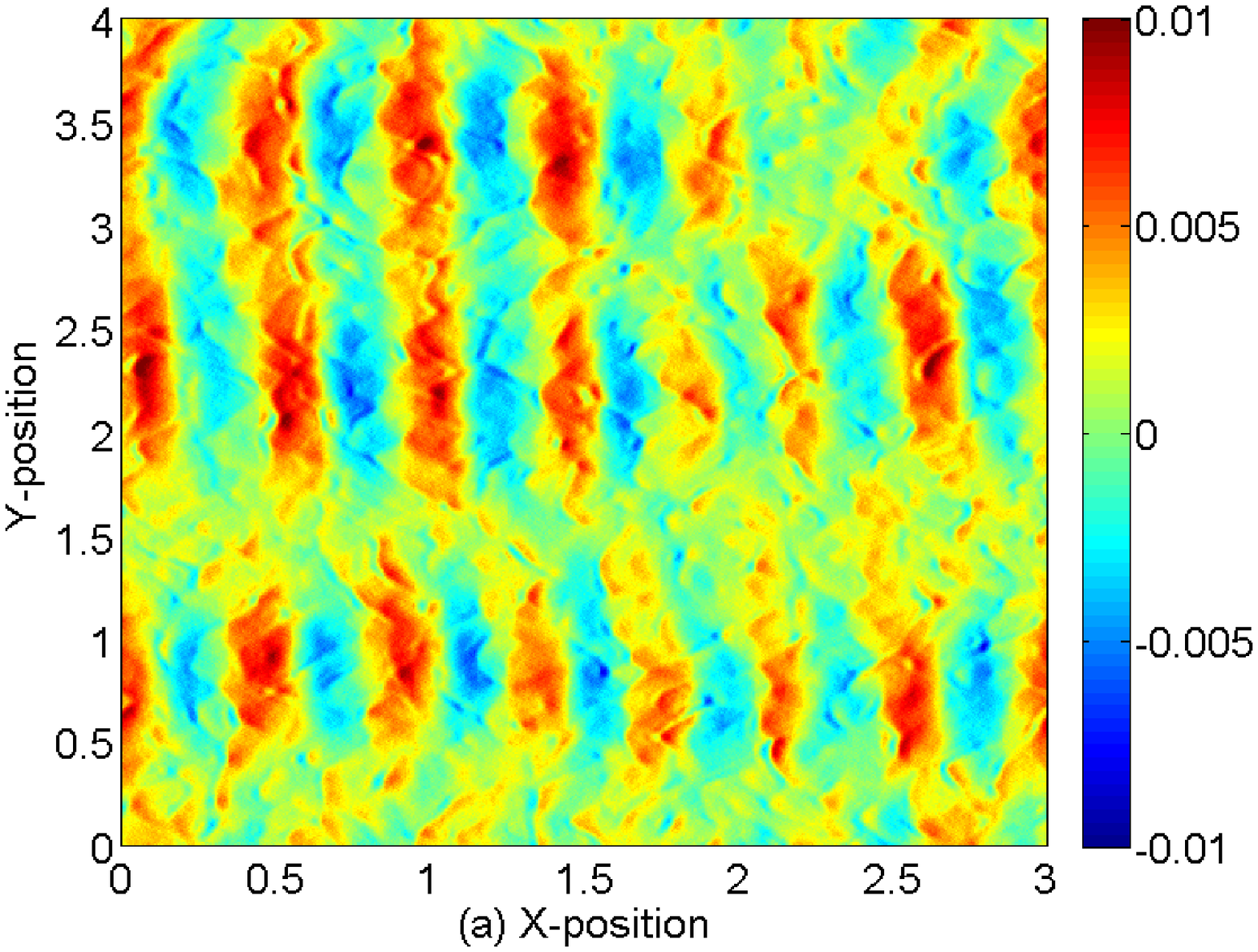}
\includegraphics[width=0.49\columnwidth]{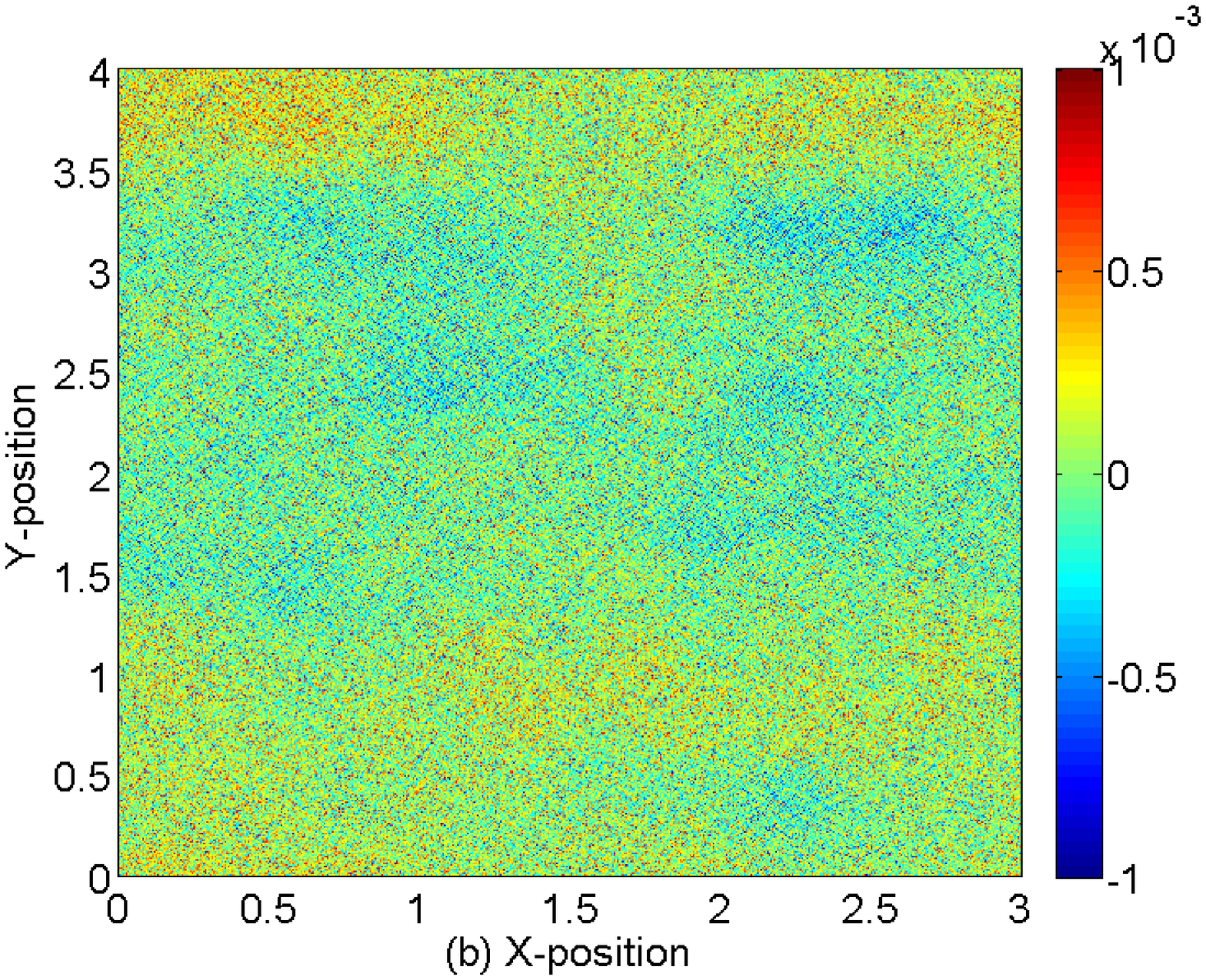}
\caption{(Color online) The electric field distributions at $t=1200$: $E_x$ is shown in 
panel (a) and $B_z$ in panel (b).}\label{Fig10}
\end{figure*}

The interaction of the electrons with the electrostatic potentials of the long waves scatters them. 
Scattering the electrons results in dissipation, which affects the current and charge density 
distribution of the plasma. Figure \ref{Fig10} shows the distributions of $E_x$ and $B_z$ at 
$T_M=1200$.
Strong electrostatic waves are still present. The magnetic field reveals strong noise 
fluctuations on a Debye-length scale, which we attribute to thermal noise. The typical 
amplitudes of $B_z$ are large; the fluctuation amplitude increases with the electron 
temperature. We conclude from Fig. \ref{Fig10}(b) that the TAWI, even if it has been 
responsible for the magnetic field growth in Fig. \ref{Fig6}, is not an efficient source 
of coherent magnetic fields for the selected parameters.

\begin{figure*}[t]
\includegraphics[width=0.49\columnwidth]{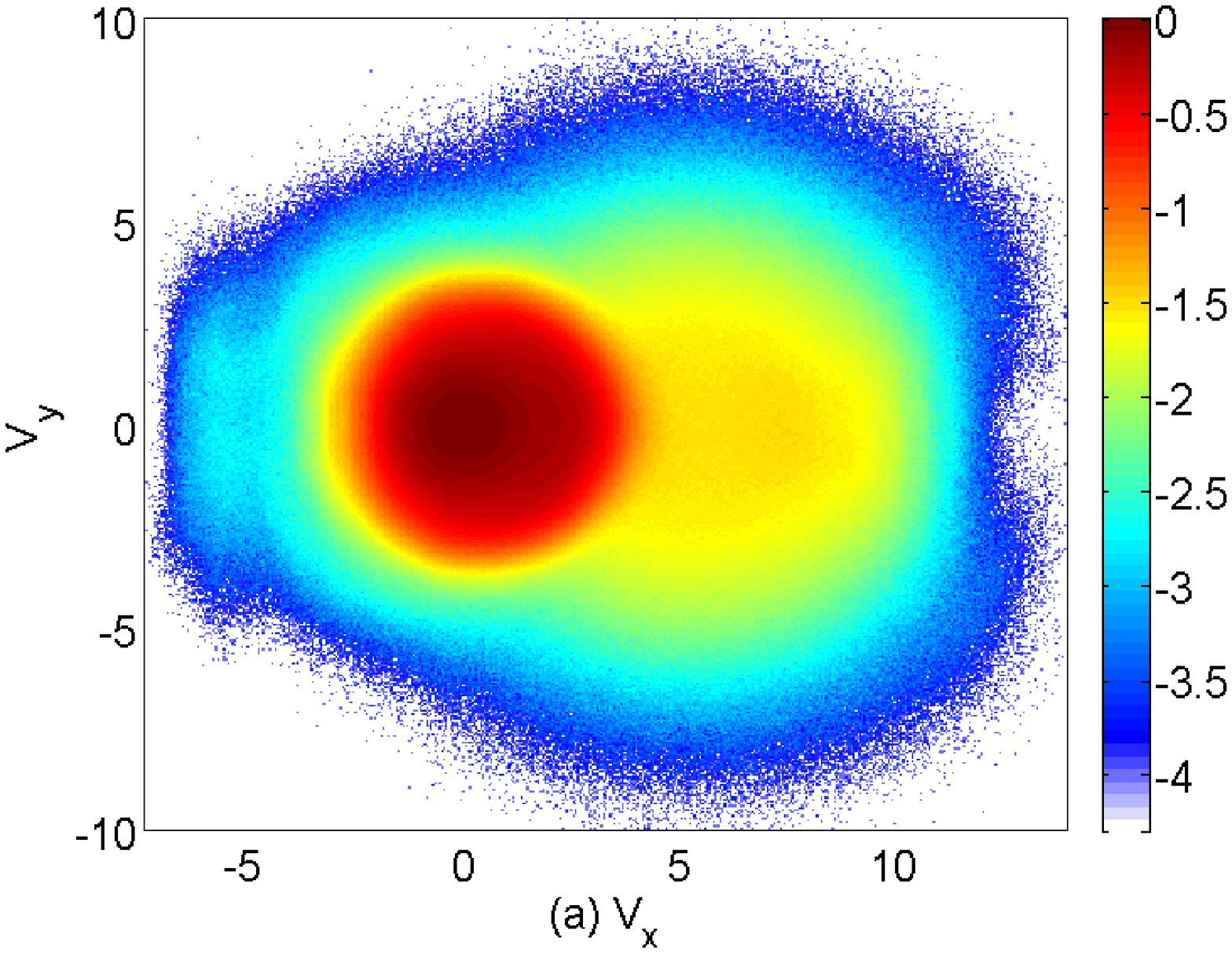}
\includegraphics[width=0.49\columnwidth]{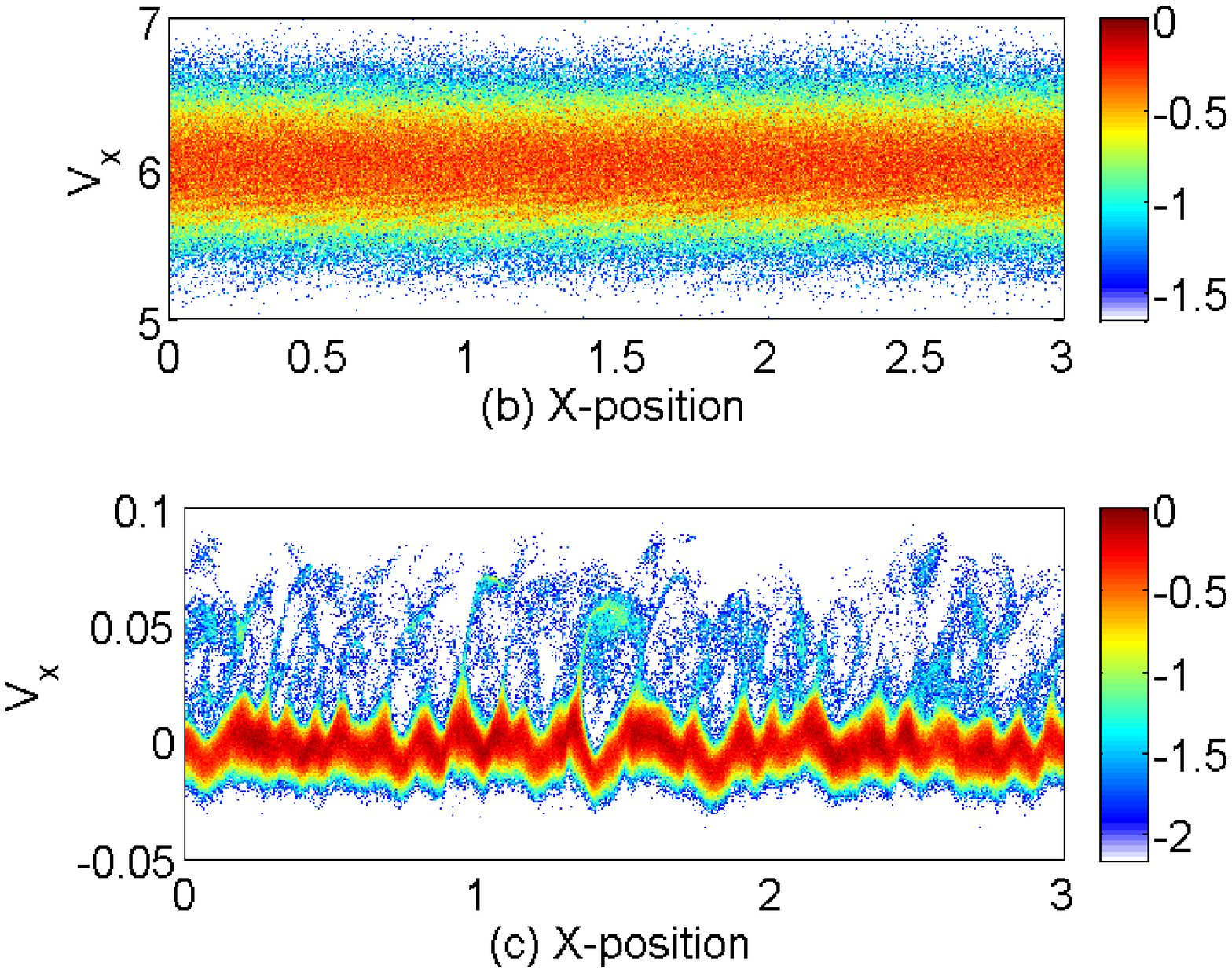}
\caption{(Color online) Particle distribution functions at $t=1200$: Panel (a) shows the electron 
velocity distribution. Panel (b) and (c) show the ion distributions integrated over the interval 
$0 < y < 0.05$. All distributions are normalized to their peak value at $t=1200$ and the colour 
scale is 10-logarithmic.}\label{Fig11}
\end{figure*}

The reason for the discrepancy between the lifetimes of the electrostatic and magnetic 
structures becomes evident from the plasma distribution functions shown in Fig. \ref{Fig11}.
The electron velocity distribution in Fig. \ref{Fig11}(a) has its peak density at $v_x \approx 0$ 
and $v_y \approx 0$. The electrons accumulate at speeds, which correspond to the phase speed of 
the waves driven by the BTI between bulk ions and electrons. An approximately circular electron 
velocity distribution extends up to $\sqrt{v_x^2+v_y^2} \approx 3$. The electron scattering is 
isotropic in the rest frame of the slow waves. A second circular structure, albeit with a lower 
number density, is observed at $v_x \approx 6$. These electrons accumulate at the phase velocity 
of the waves driven by the BTI between electrons and the fast ion beam. The electron scattering 
is apparently isotropic in the rest frame of the fast waves. The electrons are thus scattered by 
two systems of electrostatic waves that have a different wave length and that move at different 
speeds. 

The scattering should result in the rapid thermalisation of the electrons and in a dissipation 
of large-scale electron currents in the simulation plane. This dissipation explains why the
$B_z$ field in Fig. \ref{Fig10}(b) is no longer coherent over large spatial scales. The 
electrostatic fields can also be supported by modulations of the ion velocity and density and 
they are thus robust against the electron scattering. The beam ions in Fig. \ref{Fig11}(b) do 
not show any non-thermal features and the density modulation is in the linear regime. Their 
high temperature implies that the beam ions can support the large electrostatic fields 
in Fig. \ref{Fig10}(a) without developing nonlinear velocity oscillations. The bulk ions in 
Fig. \ref{Fig11}(c) show strong velocity oscillations. Their low thermal pressure implies 
that they can not sustain large electrostatic fields without showing non-thermal signatures, 
like accelerated ions.

\section{Discussion}

We have examined the growth and saturation of electrostatic Buneman-type instabilities (BTI's) 
with the help of a particle-in-cell simulation. Our system of two counter-streaming ion beams 
is representative for the foreshock region of supernova remnant shocks. The bulk ions and 
the electrons are the upstream plasma, while the ion beam corresponds to the shock-reflected 
ions. The system has initially been charge and current neutral and the plasma has been field-free, 
which is idealized. Electrostatic unmagnetized shocks have been observed experimentally and 
numerically \cite{ESshock1,ESshock2}, but their speed is typically well below the electron thermal 
speed. Their shock-reflected ion beam can thus not be faster than the electron thermal speed.
We assume here that the SNR shock carries a magnetic field perpendicular to the shock normal,
which reflects the incoming upstream ions even at high shock speeds, that is weak enough to be 
negligible with regard to the growth of the BTI. The latter implies that, in the electron rest 
frame, the electron cyclotron frequency should be well below the frequency of the waves driven 
by the BTI. A stronger magnetic field yields additional unstable wave branches \cite{MyMag} and 
isotropizes the electrons more rapidly orthogonally to the magnetic field.

The simulation has shown that two wave modes are growing. One instability branch corresponds to 
the BTI between the bulk ions and the electrons. It is the faster growing one. The second branch 
develops between the fast ion beam and the electrons. The large phase speed of its waves can 
accelerate electrons to larger speeds and this instability is thus more powerful. The electron
speeds do, however, remain nonrelativistic as in previous 1D simulation studies \cite{MyUnM}.
The BTI in unmagnetized plasma can thus not account for electron injection into the diffusive
acceleration and magnetic fields would be necessary for this purpose \cite{Amano2}. 
Both wave branches support a continuos wave spectrum ranging from the beam-aligned modes with a 
high phase speed to the slower oblique modes \cite{Amano}. The low relative speed between the 
electrons and the bulk ions implied that the instability saturated by the conventional trapping 
of electrons by beam-aligned modes \cite{H1}. The growth of magnetic fields could be observed 
after this saturation. Their energy density was, however, too low to identify unambigously the 
instability responsible for their growth. A likely candidate is the TAWI \cite{Schlickeiser}. 
The peak ratio between the electron cyclotron frequency and the plasma frequency remained below 
$10^{-3}$, which is comparable to the equivalent in the ISM.  

The electrostatic instability between the fast ion beam and the heated electrons developed at a 
later time. It saturated by the trapping of electrons by beam-aligned modes and by obliquely 
propagating slower modes. The phase space distribution revealed the simultaneous development of 
phase space holes and of the crescent distribution, which was reported first in Ref. \cite{Amano}. 
The second mechanism becomes dominant, if the beam speed exceeds by far the electron thermal 
spread, which was not the case here. The formation of such a distribution might also be aggravated 
by the much hotter ions we have used here. No growth of a strong coherent magnetic field could be 
observed, in spite of a stable thermal anisotropy of the electron distribution. 

It is unclear, why the TAWI is inefficient here. A possible reason is that the magnetic interaction
between electrons, which gives rise to the TAWI, competes with the electron interaction with the 
electrostatic wave fields. The latter has not been taken into account in numerical studies and may 
inhibit the formation of stable current filaments that are needed to sustain strong magnetic fields. We conclude 
that, at least for our simulation parameters, the TAWI driven by the BTI can not magnetize SNR shocks 
due to its short life time and low peak amplitude of the magnetic fields.

{\bf Acknowledgements:} This work was supported by Vetenskapsr\aa det (DNR 2010-4063), by 
EPSRC (EP/D043808/1, EP/D06337X/1, EP/I031766/1), by Consejeria de Educacion y Ciencia 
(ENE2009-09276), by the Leverhulme Trust (ECF-2011-383) and by the Junta de Comunidades de 
Castilla-La Mancha (PAI08-0182-3162). Computer time and support was provided by the HPC2N in 
Ume\aa.


\begin{thebibliography}{}


\bibitem{Foreshock} J P Eastwood, E A Lucek, C Mazelle, K Meziane, Y Narita, 
J Pickett, R A Treumann 2005 \textit{Space Sci. Rev.} {\bf 118} 41

\bibitem{Cargill} P J Cargill and K Papadopoulos 1988 \textit{Astrophys. J.}
{\bf 329} L29

\bibitem{ForeshockSNR} K G McClements, R O Dendy, R Bingham, J G Kirk
and L O Drury 1997 \textit{Mon. Not. R. Astron. Soc.} {\bf 291} 241

\bibitem{Schmitz} H Schmitz, S C Chapman and R O Dendy 2002 \textit{Astrophys. J.} 
{\bf 570} 637

\bibitem{Chapman} S C Chapman, R E Lee and R O Dendy 2005
\textit{Space Sci. Rev.} {\bf 121} 5

\bibitem{Umeda} T Umeda, Y Masahiro and R Yamazaki 2008 \textit{Astrophys. J.}
{\bf 681} L85

\bibitem{Matsu} S Matsukiyo and M Scholer 2006 \textit{J. Geophys. Res.} {\bf 111} 
A06104

\bibitem{Lembe} B Lembege, P Savoini, P Hellinger and P M Travnicek 2009 
\textit{J. Geophys. Res.} {\bf 114} A03217

\bibitem{Bykov} A M Bykov and R A Treumann 2011 \textit{Astron. Astrophys.
Rev.} {\bf 19} 42

\bibitem{Buneman1} O Buneman 1958 \textit{Phys. Rev. Lett.} {\bf 1} 8

\bibitem{Buneman2} O Buneman 1959 \textit{Phys. Rev.} {\bf 115} 503 

\bibitem{Dawson} J M Dawson 1983 \textit{Rev. Mod. Phys.} {\bf 55} 403

\bibitem{MyUnM} M E Dieckmann, P Ljung, A Ynnerman and K G McClements 
2000 \textit{Phys. Plasmas} {\bf 7} 5171 

\bibitem{ShimUnM} N Shimada and M Hoshino 2003 \textit{Phys. Plasmas} 
{\bf 10} 1113

\bibitem{Pavan} J Pavan, P H Yoon and T Umeda 2011 \textit{Phys. Plasmas}
{\bf 18} 042307

\bibitem{MyMag} M E Dieckmann, K G McClements, S C Chapman, R O Dendy and 
L O C Drury 2000 \textit{Astron. Astrophys.} {\bf 356} 377

\bibitem{ShimMag} N Shimada and M Hoshino 2004 \textit{Phys. Plasmas} 
{\bf 11} 1840

\bibitem{Amano} T Amano and M Hoshino 2009 \textit{Phys. Plasmas} 
{\bf 16} 102901

\bibitem{Ohira} Y Ohira and F Takahara 2007 \textit{Astrophys. J.}
{\bf 661} L171

\bibitem{DieckNJP} M E Dieckmann and A Bret 2008 \textit{New J. Phys.}
{\bf 10} 013029




\bibitem{Weibel} E S Weibel 1959 \textit{Phys. Rev. Lett.} {\bf 52} 83

\bibitem{BretW} A Bret 2007 \textit{Contrib. Plasma Phys.} {\bf 47} 113

\bibitem{Kaang} H H Kaang, C M Ryu and P H Yoon 2009 
\textit{Phys. Plasmas} {\bf 16} 082103
 
\bibitem{Stockem1} A Stockem, M E Dieckmann and R Schlickeiser
2009 \textit{Plasma Phys. Controll. Fusion} {\bf 51} 075014

\bibitem{Palodhi} L Palodhi, F Califano and F Pegoraro
2009 \textit{Plasma Phys. Controll. Fusion} {\bf 51} 125006 

\bibitem{Stockem2} A Stockem, M E Dieckmann and R Schlickeiser
2010 \textit{Plasma Phys. Controll. Fusion} {\bf 52}, 085009  

\bibitem{Lazar} M E Innocenti, M Lazar, S Markidis, G Lapenta and S Poedts
2011 \textit{Phys. Plasmas} {\bf 18} 052104

\bibitem{Tautz} R C Tautz 2011 \textit{Phys. Plasmas} {\bf 18} 012101

\bibitem{Romanov} D V Romanov, V Y Bychenkov, W Rozmus, C E Capjack and R Fedosejevs
2004 \textit{Phys. Rev. Lett.} {\bf 93} 215004

\bibitem{Pohl} J Niemiec, M Pohl, T Stroman and K Nishikawa 2008 \textit{Astrophys. J.}
{\bf 684} 1174

\bibitem{Schlickeiser} R Schlickeiser 2005 \textit{Plasma Phys. Controll. Fusion} {\bf 47} A205

\bibitem{Volk} EG Berezhko, LT Ksenofontov and HJ Volk 2003 \textit{Astron. Astrophys.} 
{\bf 412} L11

\bibitem{Cosmo} A R Bell 2004 \textit{Mon. Not. R. Astron. Soc.} {\bf 353} 550 


\bibitem{H1} K V Roberts and H L Berk 1967 \textit{Phys. Rev. Lett.} 
{\bf 19} 297

\bibitem{H2} D L Newman, M V Goldman, M Spector and F Perez 2001 \textit{Phys. Rev. Lett.} 
{\bf 86} 1239

\bibitem{H3} C Lancelotti and J J Dorning 2003 \textit{Phys. Rev. E} 
{\bf 68} 026406

\bibitem{H4} A Luque and H Schamel 2005 \textit{Phys. Rep.} {\bf 415} 261

\bibitem{H5} G Sarri \textit{et al} 2010 \textit{Phys. Plasmas} {\bf 17} 
010701 

\bibitem{H6} MY Wu, QM Lu, C Huang and S Wang 2010 \textit{J. Geophys. Res.}
{\bf 115} A10245

\bibitem{Experiment} D Prono, B Ecker, N Bergstrom and J Benford 1975
\textit{Phys. Rev. Lett.} {\bf 35} 438

\bibitem{Review} A Bret, L. Gremillet and M. E. Dieckmann 2010 \textit{Phys. Plasmas} {\bf 17} 120501

\bibitem{FluidWater} A Bret and C Deutsch 2006 \textit{Phys. Plasmas} {\bf 13} 042106



\bibitem{Eastwood} J W Eastwood 1991 \textit{Comput. Phys. Commun.} 
{\bf 64} 252


\bibitem{PhysScripta} ME Dieckmann, A Ynnerman, SC Chapman, G Rowlands and N Andersson
2004 \textit{Phys. Scripta} {\bf 69} 456

\bibitem{ESshock1} L Romagnani \textit{et al} 2008 \textit{Phys. Rev. Lett.}
{\bf 101} 025004

\bibitem{ESshock2} G Sarri, GC Murphy, ME Dieckmann, A Bret, K Quinn,
I Kourakis, M Borghesi, LOC Drury and A Ynnerman 2011 \textit{New J. Phys.}
{\bf 13} 073023

\bibitem{Amano2} T Amano and M Hoshino 2007 \textit{Astrophys. J.} {\bf 661} 190

\end{thebibliography}
\end{document}